\documentclass[useAMS,usenatbib]{mn2e}
\usepackage{latexsym}
\usepackage{makeidx}
\usepackage{cite}
\usepackage{amsmath}
\usepackage{color}
\usepackage{upgreek}
\usepackage{morefloats}
\usepackage{lineno}
\usepackage{setspace}
\usepackage{graphicx}

\title[On the potential of the EChO mission to characterise gas giant atmospheres]{On the potential of the EChO mission to characterise gas giant atmospheres}
\author[J.~K. Barstow et al.]{J.~K. Barstow$^{1,2}$\thanks{E-mail:
j.barstow1@physics.ox.ac.uk (JKB)}, S. Aigrain$^{1}$,
P.~G.~J. Irwin$^{2}$, N. Bowles$^{2}$, L.~N. Fletcher$^{2}$, J.-M. Lee$^2$\\
$^{1}$Astrophysics, Denys Wilkinson Building, University of Oxford, UK\\
$^{2}$Atmospheric, Oceanic and Planetary Physics, Clarendon Laboratory, University of Oxford, UK}
\begin{document}

\date{Submitted August 2012}

\pagerange{\pageref{firstpage}--\pageref{lastpage}} \pubyear{2012}

\maketitle

\label{firstpage}

\begin{abstract}
Space telescopes such as EChO (Exoplanet Characterisation Observatory) and JWST (James Webb Space Telescope) will be important for the future study of extrasolar planet atmospheres. Both of these missions are capable of performing high sensitivity spectroscopic measurements at moderate resolutions in the visible and infrared, which will allow the characterisation of atmospheric properties using primary and secondary transit spectroscopy. We use the NEMESIS radiative transfer and retrieval tool \citep{irwin08,lee12} to explore the potential of the proposed EChO mission to solve the retrieval problem for a range of H$_2$-He planets orbiting different stars. We find that EChO should be capable of retrieving temperature structure to $\sim$200 K precision and detecting H$_2$O, CO$_2$ and CH$_4$ from a single eclipse measurement for a hot Jupiter orbiting a Sun-like star and a hot Neptune orbiting an M star, also providing upper limits on CO and NH$_3$. We provide a table of retrieval precisions for these quantities in each test case. We expect around 30 Jupiter-sized planets to be observable by EChO; hot Neptunes orbiting M dwarfs are rarer, but we anticipate observations of at least one similar planet. 
\end{abstract}


\maketitle

\section{Introduction}
The future launches of space telescopes such as EChO (Exoplanet Characterisation Observatory, \citealt{tinetti11}) and JWST (James Webb Space Telescope) have the potential to introduce a new era in the study of extrasolar planet atmospheres. Designs for both missions incorporate the capability to perform high sensitivity spectroscopic measurements at moderate resolutions in the visible and infrared. Observations will be made of exoplanets that transit their host star from the perspective of an observer on Earth; properties of their atmospheres can then be determined by comparing the in-transit and out-of-transit fluxes of the system over a range of wavelengths, a technique called transit spectroscopy \citep{coust97,seager00}. This technique translates the small additional reduction in flux caused by the planet's atmosphere as the planet crosses the stellar disc to a measurement of atmospheric opacity at each wavelength, which in turn provides information about the atmospheric scale height, aerosols and absorbing gases in the atmosphere. When the planet is eclipsed by the star, the difference between in and out of transit fluxes at each wavelength gives the emission spectrum of the planet's dayside, which as well as providing information about absorbing gases can place constraints on the temperature structure.

Whilst missions like EChO and JWST would significantly advance our capability to perform this kind of measurement, the degree of improvement in atmospheric retrievals is highly dependent on the spectral range, resolution and signal to noise of the instrument. Solving the inverse problem \citep{rodg00} for planetary atmospheres is complex and is often hampered by insufficient information in spectra to break degeneracies in the solutions. \citet{tessenyi12} explore the number of measurements required to observe primary and secondary transit spectra with a dedicated space mission for different planet cases, but they do not consider retrievability for different atmospheric scenarios. This paper explores the potential of an EChO-like telescope to solve the retrieval problem for a range of planets orbiting different stars; the results are also applicable to any telescope with a similar spectral range (0.4---16 $\upmu$m), resolving power and noise level. 

We use the NEMESIS radiative transfer and retrieval tool developed by \citet{irwin08} to generate synthetic spectra for model giant planets under different conditions, with the expected EChO spectral range and resolution. We then add noise to the synthetic spectra and feed them back into NEMESIS to retrieve the atmospheric properties of the model planet. Comparing the retrieved atmospheric state with the original model indicates whether EChO could provide sufficient information for NEMESIS to correctly solve the retrieval problem for a real planet under similar conditions.  

NEMESIS, the Non-linear optimal Estimator for MultivariateE spectral analysis, was originally developed to analyse data from the CIRS infrared instrument on the Saturn probe, Cassini. It has since been developed to work for any solar system body, and more recently to simulate primary transit and secondary eclipse spectra for extrasolar planets \citep{lee12}.  It is a particularly suitable tool for this task because of its proven versatility, and also its efficient approach to solving the radiative transfer equation. 

NEMESIS utilises the correlated-k approximation \citep{goodyyung,lacis91} to calculate fast forward models, which speeds up the retrieval process. The correlated-k approximation allows absorption coefficients $k$ to be ranked in order of strength across a spectral interval and pre-tabulated, and relies on the assumption that absorption line strengths are well-correlated between model atmospheric layers, i.e. lines that are strongest in the lowest atmospheric layer are also strong in the layer above. This approach reduces calculation time over the line-by-line approach by reducing the number of ordinates over which the integration takes place, and the fast forward model calculation is coupled with an efficient optimal estimation scheme (e.g., \citealt{rodg00}). The user provides NEMESIS with an initial guess and an associated error, the \textit{a priori} solution, which acts to prevent overfitting and stops retrieval solutions from becoming unphysical. The best-fit solution is found by an iterative approach in which a forward model and the radiance derivatives with respect to the parameters to be retrieved are calculated at each step, and compared with the input spectrum; the next spectrum can then be calculated based on the radiance derivative and the value of a cost function, which represents the difference between the measured and synthetic spectra together with the deviation from the \textit{a priori} solution. The optimal solution is achieved when the cost function is minimised. For further details about the structure of NEMESIS and its use for retrievals of extrasolar planet atmospheres, see \citet{irwin08} and \citet{lee12}. 

\section{EChO}

At the time of writing, the proposed structure of the EChO telescope (as defined in the EChO Science Requirements and Payload Definition Documents: \citealt{echoscireq,echopdd}) is a Cassegrain telescope with a flat folding mirror and a set of dichroic mirrors for spectral separation of the light. The effective collecting area is 1.131 m$^2$ and the field of view 20''$\times$20''. The light will be split into six or seven channels, covering the wavelength range between 0.4 and 16 $\upmu$m. The visible channel detector, covering the 0.4---0.8 $\upmu$m range, is likely to be a silicon-based CCD; the short to medium infrared channels, 0.8---1.5, 1.5---2.5 and 2.5---5 $\upmu$m, will use HgCdTe detectors; the long-wave infrared channels, 5---8.5, 8.5---11 and 11---16 $\upmu$m, will probably be Si:As detectors, although other options are being investigated \citep{echopdd}. The target spectral resolving power is R$\sim$300 at wavelengths less than 5 $\upmu$m, and R$\sim$30 longwards of this. Possible descopes and trade-offs that might reduce this wavelength range and/or resolving power are not discussed here and are beyond the scope of this paper.

There are two major advantages to EChO over JWST for observations of extrasolar planets; the first is that EChO is dedicated to exoplanet observations, meaning that more time is available for revisiting interesting targets and opening up the possibility to study temporal variation in exoplanet atmospheres (for targets around quiet stars); the second is that a single instrument covers the full wavelength range between 0.4 and 16 $\upmu$m, avoiding the necessity of stitching together temporally dislocated results from different instruments to make a full spectrum. Stellar activity and instrument systematics make accurate stitching of spectra in different wavelength ranges taken at different times a complicated, sometimes impossible, process; \citet{tinetti10} discuss this problem in the context of averaging multiple observations taken within the same spectral range. JWST will launch earlier with a significantly larger mirror (an effective collecting area of 25 m$^2$ versus EChO's 1.131 m$^2$) but its exoplanet science will be limited to a few select targets, whereas EChO hopes to characterise at least 100 atmospheres.

\section{Model Planets and Synthetic Spectra}

\begin{table}
\centering
\begin{tabular}[c]{|c|c|c|}
\hline
Planet & Mass ($\times$10$^{24}$ kg) & Radius (km)\\
\hline
Hot Jupiters & 1800 & 75000 \\
Warm Jupiter & 1800 & 75000 \\
Hot Neptune & 180 & 30000 \\
\hline
\end{tabular}
\caption{Masses and radii of the model planets in this study.\label{planet_properties}}
\end{table}

In order to examine the level of information about exoplanet atmospheres that would be available from EChO, we use NEMESIS to generate a series of synthetic spectra based on a range of model planets. The parameter space that could be explored is very large, and we have reduced this to a few cases that we think best represent EChO's main targets. In this paper, we only discuss H$_2$-He gas giant planets with trace amounts of H$_2$O, CO$_2$, CO, CH$_4$ and NH$_3$, of approximately Jupiter and Neptune size. In future papers we intend to investigate effects of alkali metals, TiO/VO and haze on the visible spectra of gas giants, and extend our analysis to the super-Earth regime. All limits quoted for EChO are taken from the Payload Definition Document \citep{echopdd} unless otherwise specified.

We consider three different types of Jovian planet; a hot Jupiter-size planet, a hot Jupiter-size planet with a temperature inversion and a warm Jupiter-size planet all orbiting a Sun-like star at 35 pc. We use the Kurucz model solar spectrum\footnotemark, which has been smoothed to the EChO spectral resolution, and assume a stellar radius of 695000 km. \footnotetext{http://kurucz.harvard.edu/stars/} The value of 35 pc was chosen as it is just under half the stated distance to the faintest G-type target (75 pc), and is 15 pc greater than the distance to the brightest G-type target, so represents a distance which should be well within the capabilities of EChO. \citet{belu11} state in their study for JWST that they expect $\sim$10 transiting hot Jupiters out to 50 pc, so extrapolating to 75 pc we conclude that around 30 of these objects should be within observable range.

We also include a hot Neptune-size planet around an M-type star. At 35 pc, there is insufficient signal to observe a planet the size of Neptune around a sun-like star with a single transit. We place the M star at 6 pc, which in the baseline case for EChO is the greatest distance over which a transit around an M star is observable, with a goal of extending this to 16 pc. Fewer studies are available of the expected occurrence rate for these planets; \citet{bonfils11} estimate that 3 out of every 100 M dwarfs would have a hot Neptune in orbit, and there are around 1000 bright M dwarfs (J<10) known within 16 pc \citep{lepine11}. Despite the low occurrence rate, then, the large number of M dwarfs indicates that at least one hot Neptune should be observable by EChO, assuming a transit probability of at least 1/30. We use the Kurucz model spectrum for M5V, and assume a radius of 97995 km (equal to that of Proxima Centauri, an M5.5V star, as given in \citealt{demory09}). Primary (transmission) and secondary (eclipse) transit spectra are obtained for all of these cases. 

We fix the temperature profile and bulk H$_2$-He composition of the planet's atmosphere, but allow the abundance of five trace gas species H$_2$O, CO$_2$, CO, CH$_4$ and NH$_3$ to vary independently of each other over the concentration range 0.1---1000 ppmv. The retrievability of different temperature profiles is implicitly tested by considering the different planet cases. The bulk atmosphere composition is assumed to be 90\% H$_2$ and 9.9\% He; \citet{lee12} found that variation in the He/H$_2$ ratio had a negligible effect on the retrieval of temperature and molecular abundances for HD 189733b, which is confirmed in the present study. We compare retrieved temperature and gas abundances for hot Jupiter model spectra with 90\% H$_2$ and 9.9\% He, retrieved firstly assuming the correct H$_2$/He ratio and secondly assuming 80\% H$_2$ and 19.9\% He, and find negligible differences. 

Very few assumptions have been made about the expected levels of trace gases in each of these atmospheres, as the objective of this study is to investigate whether features due to these gases, if present, would be detectable in each case. Initially, we assume a constant volume mixing ratio (VMR) as a function of altitude for all gases, which is the simplest possible scenario. In Section~\ref{altvar} we take a brief look at a more `realistic' vertical profile for CH$_4$ and examine how our findings might be affected by changes in the VMR as a function of altitude. 

We generate between 100 and 500 synthetic transit and eclipse spectra for each test scenario, each with different uniformly distributed trace gas VMRs. We then retrieve the VMRs (and temperature as a function of atmospheric pressure in the eclipse case) from each synthetic. We calculate a reduced $\chi^2$ parameter for each retrieval, which represents the goodness of fit within the measurement error and is the $\chi^2$ divided by the number of degrees of freedom in the retrieval\footnotemark. This value should be $\sim$1 for a good retrieval. 

\footnotetext{The number of degrees of freedom is given by $n_{\mathrm{measurements}}-n_{\mathrm{parameters}}-1$. When retrieving a continuous profile with some correlation length, it is difficult to exactly define the number of parameters. Here, we simply use the number of atmospheric levels for a continuous retrieval, plus the number of gas scaling factors retrieved. There are fewer parameters than this in reality, so our estimates of $\chi^2/n$ are conservatively large, but since the number of measurements (605) far outweighs the maximum number of parameters (55) we do not severely overestimate the $\chi^2$, and it is clear that even with a slight overestimation we can achieve $\chi^2/n$ values of order 1.}

\subsection{Gaseous absorption}
\begin{table}
\centering
\begin{tabular}[c]{|c|c|}
\hline
Gas & Source\\
\hline
H$_2$O & HITEMP2010 \citep{roth10}\\
CO$_2$ &  CDSD-1000 \citep{tash03}\\
CO & HITEMP1995 \citep{roth95}\\
CH$_4$ & STDS \citep{weng98}\\
NH$_3$ & HITRAN2008 \citep{roth09}\\
\hline
\end{tabular}
\caption{Sources of gas absorption line data.\label{linedata}}
\end{table}

Line strength information for H$_2$O, CO$_2$, CO and CH$_4$ and collision-induced absorption for H$_2$-H$_2$ and H$_2$-He are all as in \citet{lee12} and references therein. In addition, we include NH$_3$ with line strengths taken from the HITRAN2008 database \citep{roth09}. The limited information available for some of these gases, particularly our omission of high-temperature line information for NH$_3$, may result in overly pessimistic predictions about retrievability of some constituents; in the future, we hope to exploit new information from sources such as \citet{tenn12} to improve the completeness of gas absorption information in our radiative transfer models, which will in turn increase the reliability of our predictions. The sources of line data for each of the five trace absorbers are summarized in Table~\ref{linedata}.

\subsection{Spectral sensitivity}

\begin{figure*}
\centering
\includegraphics[width=0.8\textwidth]{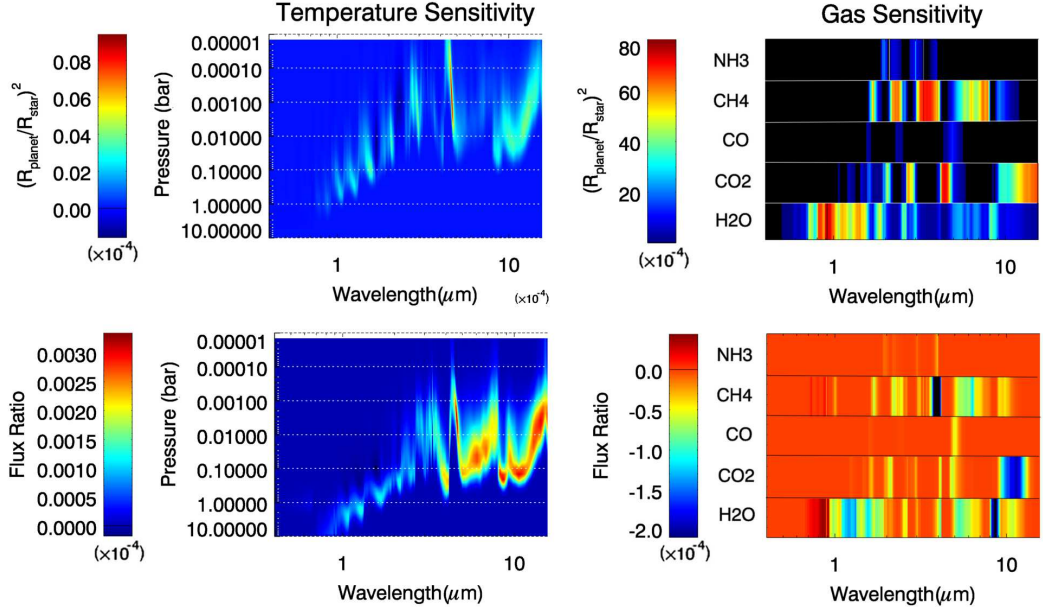}
\caption{Sensitivity of a hot Jupiter planet to various atmospheric
  parameters. The top plots show the derivative of the squared radius
  ratio with respect to a 1K change in temperature at each altitude
  level (left) and with respect to a multiplication of the whole
  vertical profile of each gas (right). Similar plots are shown below
  for the change in flux ratio in eclipse.\label{sensitivities}}
\end{figure*}

The level of sensitivity to different atmospheric properties in transmission and eclipse (secondary transit) geometries is indicated by the derivative plots in Figure~\ref{sensitivities}. It can be seen that there is equal sensitivity to temperature across a wide range of altitudes for the eclipse case when the whole spectrum is considered, but the sensitivity to temperature in transmission is heavily skewed towards the atmosphere above the 10-mbar level, because the atmosphere is opaque out to higher altitudes in transmission. This means that there is insufficient information to retrieve temperature as a function of pressure in transmission geometry. There is very little sensitivity to CO and NH$_3$ in either geometry, but there are several spectral regions in which H$_2$O, CO$_4$ and CH$_4$ have a measurable effect on the spectrum. The narrow regions of sensitivity to CO and NH$_3$, at 5 and 4 $\upmu$m respectively, are heavily contaminated by absorption lines of the other gases, producing degeneracies in the solution. A key region of the spectrum for CO$_2$ is the 15 $\upmu$m band; maximum sensitivity occurs at the centre of this band in transmission geometry, and in the wing of the band in eclipse.

\subsection{Noise Calculation}
\label{noisecalc}
We assume that the instrument is entirely photon-limited, and estimate the noise based on the square root of the number of photons per channel during the transit. Whilst we may expect noise to be as much 2---3$\times$ photon-limited, this could easily be reduced to the photon limit by averaging over a greater number of transits, provided the noise is uncorrelated.  Star spots are likely to be problematic for targets orbiting M stars as these are often active, and changing configurations of spots could present difficulties when averaging over multiple transits. The presence of correlated noise and systematic noise related to the instrument or star would be a severe problem, but the effect is difficult to quantify for EChO at this stage of the design and is beyond the scope of this paper. The equations used to calculate the noise per channel are given below.

\begin{equation}
n_{\lambda}=I_{\lambda}{\,}{\pi}{\,}(r_{\star}/D_{\star})^2{\,}({\lambda}/hc){\,}({\lambda}/R){\,}A_{\mathrm{eff}}{\,}QE{\,}{\eta}{\,}t\label{eqn1}
\end{equation}
where $n_{\lambda}$ is the number of photons received for a given wavelength $\lambda$, $I_{\lambda}$ is the spectral radiance of the stellar signal, $r_{\star}$ is the stellar radius, $D_{\star}$ is the distance to the star, $h$ and $c$ are the Planck constant and speed of light, $R$ is the spectral resolving power, $A_{\mathrm{eff}}$ is the telescope effective area, $QE$ is the detector quantum efficiency, $\eta$ is the the throughput and $t$ is the exposure time. The resolving power, effective area, quantum efficiency (assumed 0.7 for all channels) and throughput per channel are all taken from the EChO Payload Definition Document \citep{echopdd}. The estimated throughput $\eta$ of each channel is given as 0.191 (0.4---0.8 $\upmu$m), 0.284 (0.8---1.5 $\upmu$m), 0.278 (1.5---2.5 $\upmu$m), 0.378 (2.5---5 $\upmu$m), 0.418 (5---8.5 $\upmu$m), 0.418 (8.5---11 $\upmu$m), 0.326 (11---16 $\upmu$m). The effective telescope area is 1.131 m$^2$. The exposure time is estimated based on a typical transit duration of 3 hours for the Sun-like star and 1 hour for the M dwarf, with an 80\% duty cycle. 

To calculate the noise on the stellar signal, the square root of the number of photons is taken and then Equation~\ref{eqn1} is reversed to obtain this in terms of spectral radiance, which gives the absolute noise on the observed radiance. We use the differential chain rule to calculate the noise in the ratio of in/out of transit fluxes, and assume that the planet signal is sufficiently small that $I_{\mathrm{in}}{\sim}I_{\mathrm{out}}$, where $I$ is the radiance. The error in the flux ratio is then given by Equation~\ref{err_ratio}:

%Equation~\ref{eqn2} below is then used to calculate the noise in the ratio of the in/out of transit fluxes, which is just a statement of the differential chain rule for error propagation.

%If $C$ is some function of $A$ and $B$ then 

%\begin{equation}
%(\sigma_C)^2=\left(\frac{{\partial}C}{{\partial}A}\sigma_A\right)^2+\left(\frac{{\partial}C}{{\partial}B}\sigma_B\right)^2\label{eqn2}
%\end{equation}

%In this case, the flux ratio $C$=$A$/$B$ is of order 1, $A{\simeq}B$ are the in/out of transit fluxes and $\sigma_A{\simeq}\sigma_B$, so the error $\sigma_C$ is given by:

%\begin{equation}
%\sigma_C=\sqrt{2\left(\frac{\sigma_A}{A}\right)^2}\label{eqn3}
%\end{equation}

\begin{equation}
\sigma_{\mathrm{ratio}}=\sqrt{2\left(\frac{\sigma_{\mathrm{I}}}{I}\right)^2}\label{err_ratio}
\end{equation}

In the transmission case, the radius ratio $(R_{\mathrm{planet}}/R_{\star})^2=1-I_{\mathrm{in}}/I_{\mathrm{out}}$. In the eclipse case, the planet-star flux ratio $F_{\mathrm{planet}}/F_{\star}=I_{\mathrm{out}}/I_{\mathrm{in}}-1$. It can be seen from both of these equations that the error in the in/out of transit radiance ratio is also the error in the radius squared and flux ratios for the two transit cases. 

The error calculated in this way is taken to be the 1$\sigma$ value for Gaussian noise on the spectrum, so the final error added to each band in the synthetic spectrum is calculated by generating a series of Gaussian-distributed random numbers with a mean of zero and $\sigma$ equal to the calculated noise. 

The scenario described above is of course a perfect scenario; whilst the aim for a telescope like EChO is to approach the photon noise limit as closely as possible, the reality is likely to include red noise and instrument systematics. \citet{gibson11} compare the error bars on a HST/NICMOS transmission spectrum obtained by fully taking into account instrument systematics using Bayesian methods with work assuming white noise only, and find that the errors are inflated by a factor of $\sim$3. This is an extremely conservative estimate, and since NICMOS systematics were particularly severe we may assume that the same increase in error for EChO represents a worst-case scenario.

\section{Planet Test Cases}
\subsection{Hot Jupiter}
\label{hotjupsection}

\begin{figure}
\centering
\includegraphics[width=0.4\textwidth]{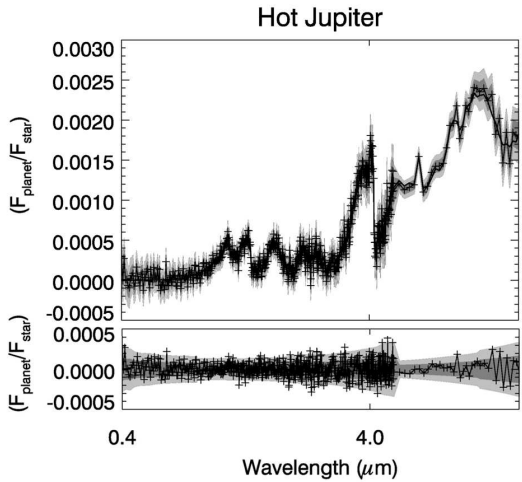}
\caption{An example of a fit to a noisy synthetic hot Jupiter eclipse spectrum. The residual is shown in the lower panel. The 1-$\sigma$ error bar is shown in dark gray shading, and the 2-$\sigma$ error in light gray. The joined crosses in the top panel show the noisy synthetic, and the solid black line shows the fitted spectrum.\label{hot_j_examplespec}}
\end{figure}

The first planet we consider is a hot Jupiter-size planet orbiting a sun-like star. This planet is similar to HD 189733b, which has been observed using the Hubble Space Telescope; evidence for H$_2$O, CO$_2$, CO and CH$_4$ in the atmosphere of HD 189733b has been presented by \citet{swain09}, \citet{madhu09} and \citet{lee12}. NH$_3$ has also been included as a possible constituent in this study as it has been detected in $\sim$1000 K brown dwarf atmospheres \citep{saumon00} which may have similarities with hot exoplanet atmospheres. To demonstrate sensitivity to different vertical temperature profiles, we include two variants of the hot Jupiter case with and without a temperature inversion above the tropopause; the case with a temperature inversion is similar to HD 209458b \citep{desert09}. Initially, we assume that the atmosphere is well-mixed and the composition is constant as a function of altitude. 

The scale height of the atmosphere is calculated based on the mass and radius provided (see Table~\ref{planet_properties}), from which we obtain the gravitational acceleration at the `surface' of the planet. In the case of a gas planet, we assume that this `surface' lies at a pressure level of 10 bar, as the atmosphere is opaque at pressures greater than this. The atmospheric scale height for the hot Jupiter varies as a function of temperature, but is of order a few hundred km. 

The initial test incorporates 500 secondary eclipse spectra with different trace gas abundances. We utilise the whole of the anticipated EChO spectral range, and assume that measurements were made during a single transit. An example noisy synthetic spectrum, with best-fit model, is shown in Figure~\ref{hot_j_examplespec}. 

\begin{figure*}
\centering
\includegraphics[width=1.0\textwidth]{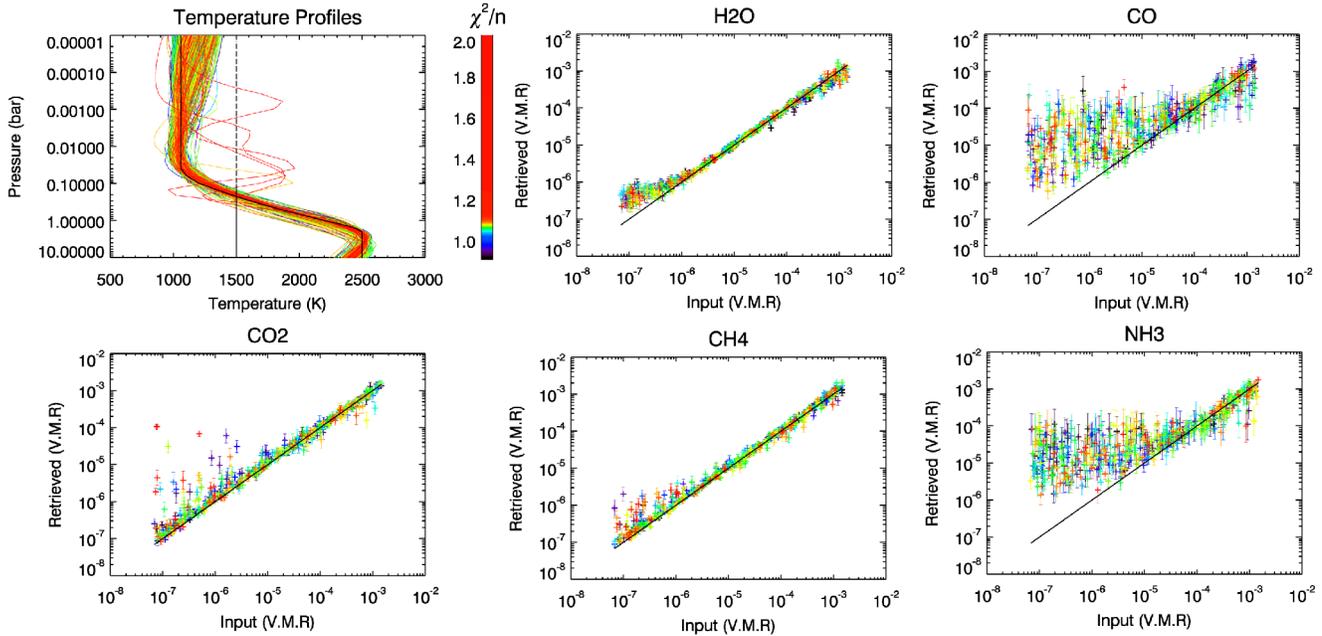}
\caption{Eclipse retrieval results for a hot Jupiter without a temperature inversion. colours correspond to different $\chi^2$ values, increasing from purple to red. The dashed black line on the temperature plot is the \textit{a priori} temperature provided to NEMESIS; the solid black line is the input temperature profile. The black lines on the VMR plots are 1:1 correspondence lines. Some of the temperature retrievals with the very highest $\chi^2$ are poor, indicating that in these few cases NEMESIS has failed to adequately fit the spectrum.\label{hotjupresults}}
\end{figure*}

\begin{figure*}
\centering
\includegraphics[width=1.0\textwidth]{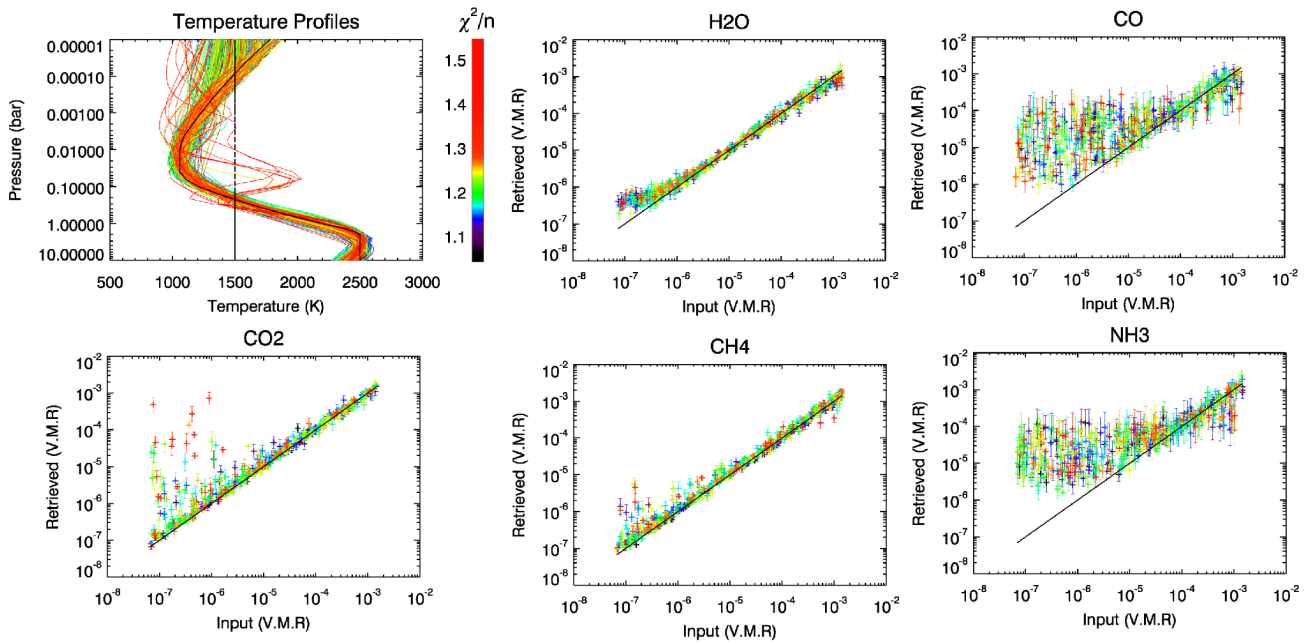}
\caption{Eclipse retrieval results for a hot Jupiter with a temperature inversion; colours/linestyles as Figure~\ref{hotjupresults}.\label{hotjupinvresults}}
\end{figure*}

We retrieve temperature as a function of pressure and altitude-independent volume mixing ratios for each of the five trace absorbers. The results are presented in Figures~\ref{hotjupresults}---\ref{hotjupinvresults}. The rainbow colours of each line/point corresponds to the reduced $\chi^2$ ($\chi^2$ divided by the number of degrees of freedom $n$) of the fit in each case, with purple points having the lowest $\chi^2/n$ and red points the highest. All but a very few retrievals have a $\chi^2/n$ below 1.2. It can be seen that in general NEMESIS can successfully retrieve the temperature profile and VMRs of H$_2$O, CO$_2$ and CH$_4$ in all cases; it is however not capable of retrieving CO and NH$_3$ where the VMR of these gases falls below $10^{-4}$, so only a rather high upper limit would be achievable for these gases. The retrieval of H$_2$O VMRs starts to become unreliable for VMRs below 5$\times10^{-7}$. 

\begin{figure}
\centering
\includegraphics[width=0.4\textwidth]{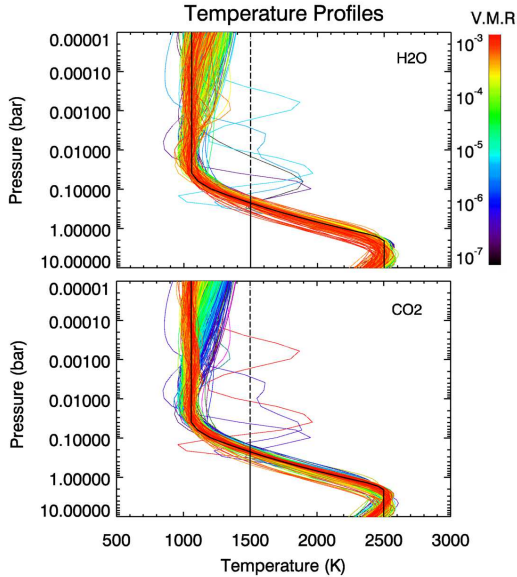}
\caption{Retrieval results for temperature; colours in the upper and lower plots correspond to different input VMRs of H$_2$O and CO$_2$ respectively, with low to high values going from purple to red. Linestyles are as in Figure~\ref{hotjupresults}.\label{temp_degeneracy}}
\end{figure}

The quality of the temperature retrieval is good in the troposphere and stratosphere, but at higher altitudes and in the very deep atmosphere results are comparatively poor. This is due to degeneracies between the temperature retrieval and the retrieval of H$_2$O and CO$_2$ abundances; when retrieving several quantities, there are sometimes multiple atmospheric scenarios that result in spectra with an equally good fit to the data, so the retrieval can converge on an incorrect solution. This is demonstrated in Figure~\ref{temp_degeneracy}; low VMRs of H$_2$O and CO$_2$ result in the retrieval of a higher mesospheric temperature than the `real' temperature. Information about temperature at a particular altitude is obtained from fitting the shape of a H$_2$O or CO$_2$ absorption feature, so a lower abundance of either gas, corresponding to a smaller feature, may result in a reduction of available information from the feature. This could mean that the retrieved temperature relaxes back towards the \textit{a priori}, which in this case is higher than the `real' temperature. 

It can be seen in these plots that in cases where the retrieved value is far from the input the error bars can be small, e.g. for small VMRs of CO$_2$. This is because NEMESIS estimates the error on each retrieved parameter based only on the errors in the measured spectrum and in the \textit{a priori}, which does not account for degeneracy. It is therefore crucial to explore the possibility of degenerate solutions by varying the \textit{a priori} and performing the retrieval multiple times, which ensures that the interdependence of solutions for different variables is understood. \citet{lee12} do this, and they also examine the variation in retrieved temperature as a function of $\chi^2/n$ for different gas VMRs. This provides a more robust estimate of the error in the retrieved temperature.

\begin{figure}
\centering
\includegraphics[width=0.4\textwidth]{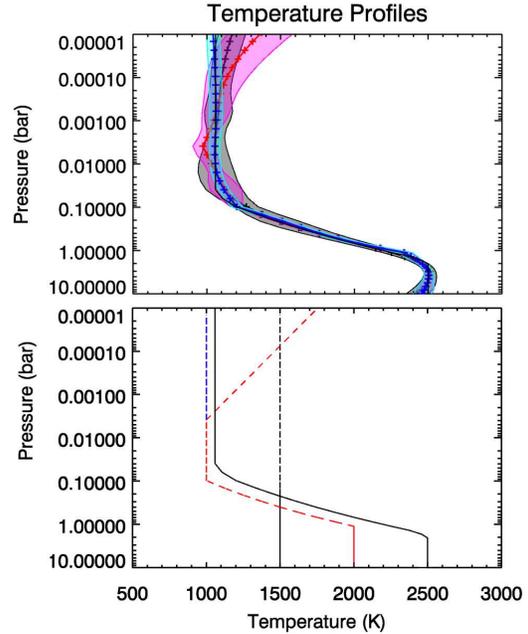}
\caption{Comparison of temperature retrievals from three different
  \textit{a priori} starting points (top panel). The mean and standard deviation over 100 retrievals are shown for each case; the joined crosses indicate mean values and the shaded envelope the standard deviation. The \textit{a priori} temperatures are shown on the bottom. The input temperature profile is the overplotted black solid line in both panels.  \label{tempaprioricompare}}
\end{figure}

As well as exploring degeneracies, it is also important to test the effect of different \textit{a priori} assumptions on the retrieved temperature profile. This test indicates the regions of the atmosphere for which information is available in the spectrum, as outside these regions the retrieved profile relaxes back to the \textit{a priori}. We test this by retrieving the first 100 spectra from the previous test, but assuming three different \textit{a priori} temperature profiles. The results of this test are shown in Figure~\ref{tempaprioricompare}. It is clear that the standard deviation in retrieved temperatures decreases as the \textit{a priori} becomes closer to the correct solution. The mean retrieved temperature is correct for all three cases below the tropopause and is close to the correct value at pressures greater than 1 mbar; the mean retrieved temperature is correct everywhere for the case where the shape of the \textit{a priori} profile is the same as the shape of the input profile. For the temperature at altitudes above the 1-mbar level there is a lack of information available in the spectrum, so the retrieval coverges on the \textit{a priori} value. From this test, we see that a better \textit{a priori} guess does slightly improve the accuracy of the retrieved temperature, but a comparatively poor \textit{a priori} assumption can still produce a valid solution provided there is information in the spectrum. The $\chi^2/n$ is very similar for all \textit{a priori} cases, with a slight reduction in the average from 1.08733 for the isothermal \textit{a priori} to 1.05709 for the adiabatic \textit{a priori}, so the choice of \textit{a priori} does not affect the quality of fit to the spectrum. In regions of low information content, the retrieved temperature profile will be affected by the \textit{a priori} and therefore it is important to perform retrievals from a range of different \textit{a priori} starting points to ensure that this behaviour is fully understood.

Similar tests for 100 spectra have been performed in transmission geometry; these tests are complicated by the fact that there is insufficient sensitivity to temperature in transmission spectra to disentangle the effects from those due to gaseous absorption, so in order to perform a retrieval one must assume a temperature structure. We test the effect of incorrect assumptions about the T-P profile on the retrievability of the gas VMRs. First, we assume the same constant temperature profile as the \textit{a priori} for the eclipse case. It is reasonable to assume a constant temperature that is within the range of temperatures in the `real' profile, as it is possible to compute an equilibrium temperature (Equation~\ref{teq}) for an exoplanet based on measurement of its orbital parameters: 

\begin{equation}
{T_{\mathrm{eq}}^4}=(1-A)\frac{R_{\star}^2T_{\star}^4}{4a^2}\label{teq}
\end{equation}

where $\star$ indicates stellar radius $R$ and temperature $T$, $A$ is the Bond albedo of the planet and $a$ is the orbital semi-major axis. This is often a poor approximation for giant planet temperatures in our own solar system, as they also have large internal heat sources, but is the best first-order approximation possible for extrasolar planets. 

Secondly, we perform the retrieval with the temperature profile from the forward model that we used to create the noisy synthetic. Since we could use a temperature constraint from eclipse measurements as an input for a transmission retrieval, which with perfect heat redistribution in the atmosphere would be the same as the temperature at the terminator, it is conceivable we could approach this level of knowledge about the terminator temperature; the level of prior knowledge about temperature structure for the majority of cases is likely to fall somewhere between these extrema. 

\begin{figure*}
\centering
\includegraphics[width=1.0\textwidth]{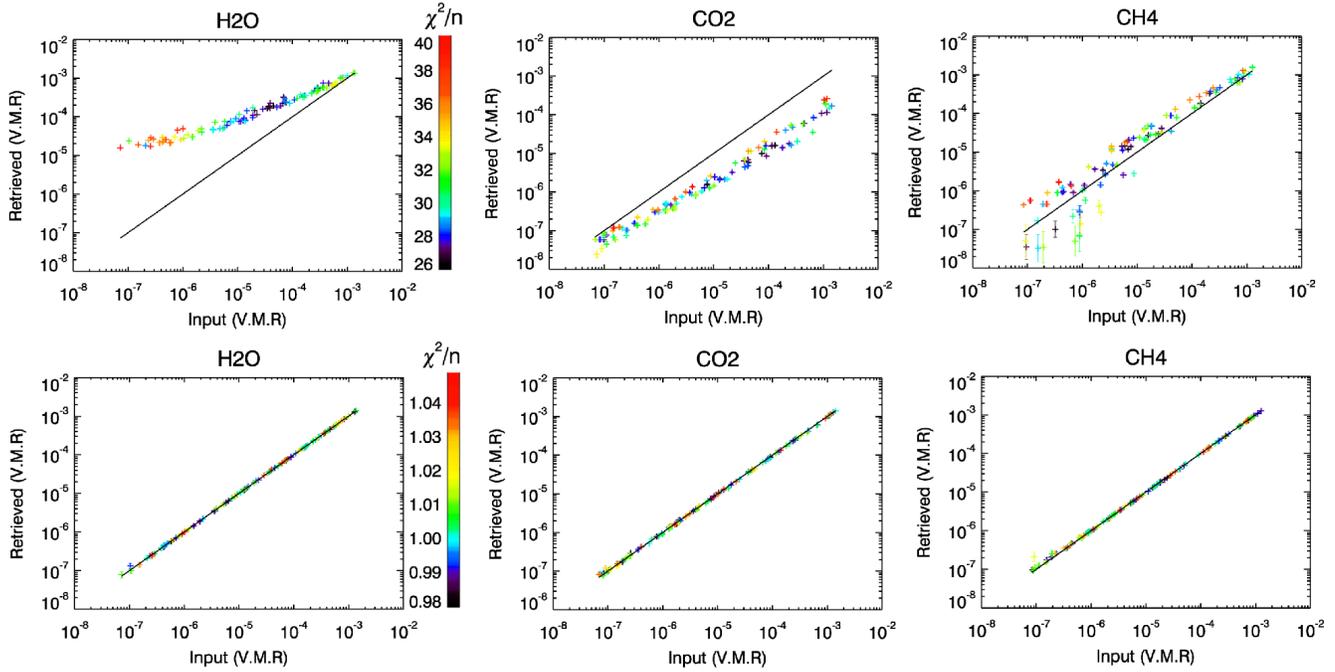}
\caption{Retrievals of H$_2$O, CO$_2$ and CH$_4$ from transmission spectra for a hot Jupiter with no temperature inversion. In the upper panels, a constant temperature of 1500 K is assumed; in the lower panels, the input temperature profile is assumed. colours are as Figure~\ref{hotjupresults}.\label{hotjupgasprimary}}
\end{figure*}

\begin{figure*}
\centering
\includegraphics[width=1.0\textwidth]{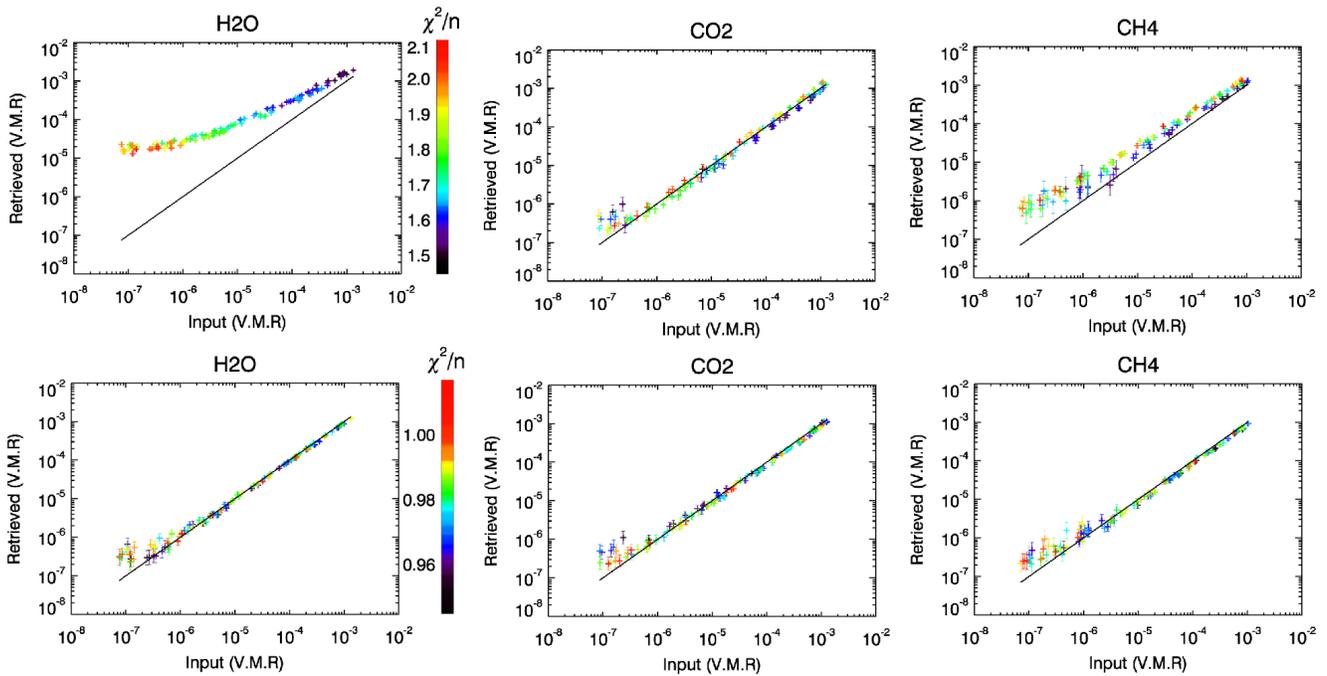}
\caption{As Figure~\ref{hotjupgasprimary}, but for a hot Jupiter with a temperature inversion.\label{hotjupgasinvprimary}}
\end{figure*}

\begin{figure*}
\centering
\includegraphics[width=0.85\textwidth]{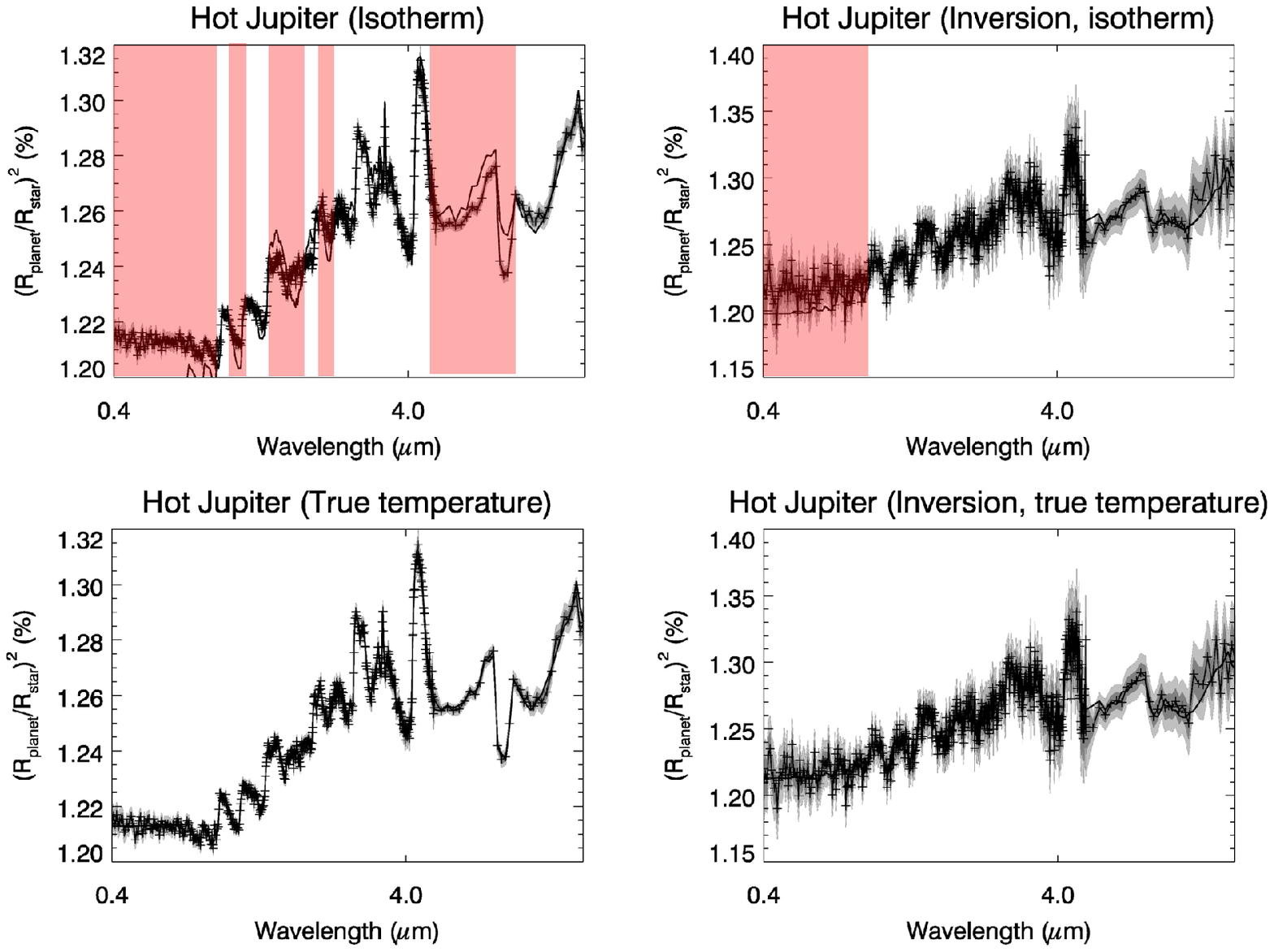}
\caption{Examples of fits to noisy synthetic hot Jupiter transmission spectra. The 1-$\sigma$ error bar is shown in dark gray shading, and the 2-$\sigma$ error in light gray. The joined crosses in the top panel show the noisy synthetic, and the solid black line shows the fitted spectrum. The left-hand panels are the hot Jupiter spectra without a temperature inversion (30 transits) and the right show the Jupiter with a temperature inversion (single observation). The red shaded areas indicate the regions where the spectral fit is poor when an isothermal temperature approximation is used. \label{hot_j_examplespec_pt}}
\end{figure*}

The results for H$_2$O, CO$_2$ and CH$_4$ in these two cases are shown in Figures~\ref{hotjupgasprimary} and~\ref{hotjupgasinvprimary}; results for CO and NH$_3$ are as poor as the results obtained from eclipse spectra. The results presented for the no-inversion case have assumed a noise level consistent with combining measurements from 30 transits to increase the signal to noise, whilst noise for a single transit has been added to the inversion case. Adding together 30 transits improves the retrieval accuracy for cases where the VMRs are low, provided the assumed temperature profile is close to reality, but no improvement is observed for the constant temperature profile. However, the difference in $\chi^2/n$ between the correct and incorrect temperature profile assumptions is much greater when the measurement error is smaller, so improving signal to noise increases the likelihood of being able to use the $\chi^2$ to distinguish between different temperature scenarios. This can be seen in Figure~\ref{hot_j_examplespec_pt}; the fit is poor in both cases (30 observations and 1 observation) where the temperature is assumed to be isothermal, but the difference in goodness of fit is much easier to discern in the left hand panels, where the signal to noise is higher.


As can be seen from these plots, an assumption of constant temperature results in a very poor retrieval of H$_2$O VMR and significantly reduces the retrieval accuracy of CO$_2$ and CH$_4$. However, the average $\chi^2/n$ for the constant temperature case is 30, whereas it is close to 1 when the correct temperature profile is used. Although the temperature structure cannot be directly retrieved from a transmission spectrum, it may therefore still be possible to place some constraint on it by searching for a minimum $\chi^2$ within a range of different assumed temperature profiles. This could be done by repeating the retrieval over a range of different temperature assumptions and choosing the solution with minimum $\chi^2$.

Even if we have no temperature information from a secondary eclipse measurement, we can still do better than the isothermal assumption by calculating the expected equilibrium temperature $T_{\mathrm{eq}}$ of the planet and modelling the atmosphere as a single slab. This slab of atmosphere radiates equally out to space and down into the lower atmosphere at the stratospheric temperature $T_{\mathrm{strat}}$. If we equate the total radiation absorbed and emitted by the slab, we can relate these two temperatures using Equation~\ref{radeq1}
\begin{equation}
e{\sigma}{T_{\mathrm{eq}}^4}=2e{\sigma}{T_{\mathrm{strat}}}^4\label{radeq1}
\end{equation}
where $e$ is the emissivity of the slab, $\sigma$ is the Stefan-Boltzmann constant and the factor of 2 arises from the fact that the slab receives radiation from space only but emits radiation equally in both directions. Rearranged, this gives the relation that $T_{\mathrm{eq}}=2^{1/4}T_{\mathrm{strat}}$.

We can then calculate the temperature at the bottom of the troposphere/surface of the planet by calculating the dry adiabatic lapse rate of the atmosphere, given by Equation~\ref{radeq2}; $g$ is the acceleration due to gravity and $c_p$ the specific heat capacity. This is to some extent arbitrary as we do not know \textit{a priori} the pressure level at which this occurs, but for a hot gaseous planet we assume that this occurs at approximately the 1 bar level; the retrieval is insensitive to temperature variations below this level (see Figure~\ref{sensitivities}). The top of the tropopause we assume to occur at around 0.1 bar, based on observations of solar system giant planets. We can assume that pressure and altitude are related by the hydrostatic equation,~\ref{radeq3}.
\begin{equation}
\Gamma=-\frac{\mathrm{d}T}{\mathrm{d}z}=-\frac{g}{c_p}\label{radeq2}
\end{equation}
\begin{equation}
p=p_0\mathrm{e}^{-\frac{z}{H}}\label{radeq3}
\end{equation}

$H$ is the atmospheric scale height, which is given by $H=kT/mg$ where $k$ is the Boltzmann constant, $T$ is temperature and $m$ is the molecular mass of the atmosphere. If we know the bulk atmospheric composition and assume for the purpose of calculating the scale height that $T$ is the stratospheric temperature, we can combine equations~\ref{radeq2} and~\ref{radeq3} to calculate the temperature at the base of the troposphere. 
\begin{equation}
T_{\mathrm{trop}}=T_{\mathrm{strat}} -\Gamma\frac{kT_{\mathrm{strat}}}{mg}\mathrm{ln}\left(\frac{p_1}{p_2}\right) \label{radeq4}
\end{equation}
Using this equation, we can derive a simple temperature profile based on the planet's equilibrium temperature.

\begin{figure*}
\centering
\includegraphics[width=1.0\textwidth]{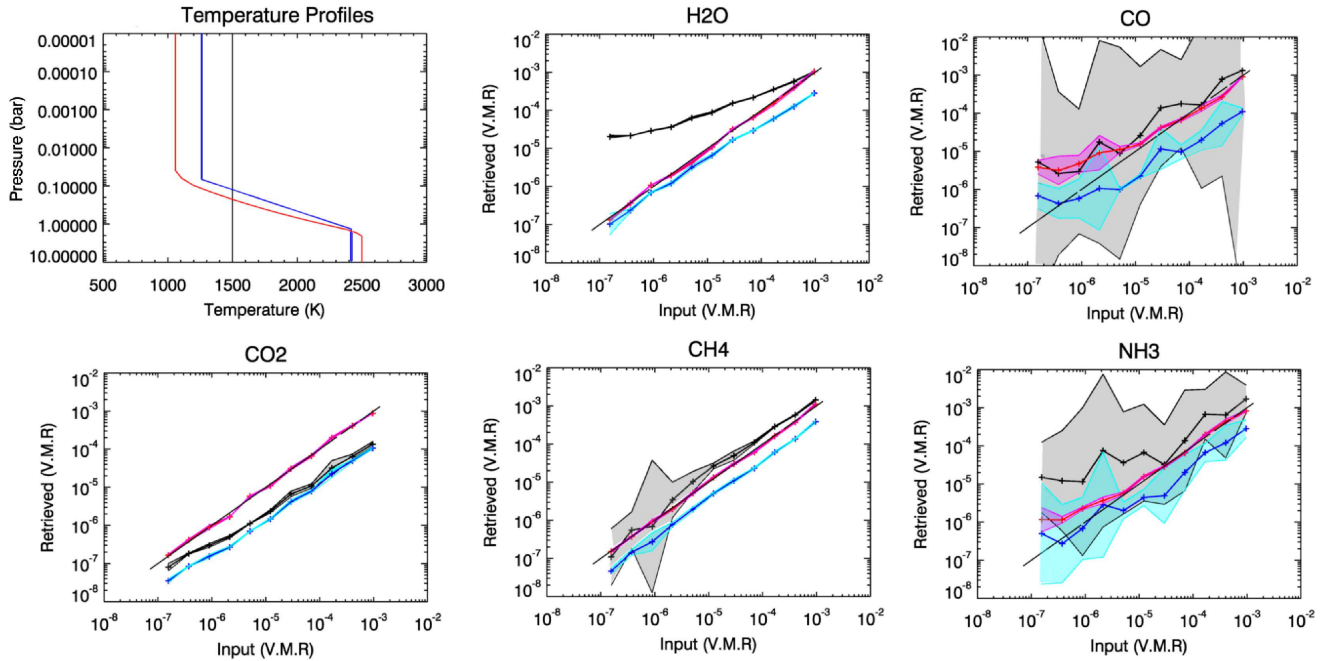}
\caption{A comparison between transmission retrievals for the hot Jupiter case where different temperature profiles (shown in the top left) are assumed. The joined crosses are the mean values and shaded areas represent the standard deviation over 100 retrievals.\label{hotjadiabattest}}
\end{figure*}

We compare the results from all three temperature profile parameterisations for a hot Jupiter without a temperature inversion in Figure~\ref{hotjadiabattest}. Even though the shape of the original temperature profile is better reproduced by the single slab atmosphere model, it does not significantly improve the retrieval result over the isothermal assumption case. The H$_2$O retrieval is improved when the estimated temperature structure is more realistic, but the CO$_2$ and CH$_4$ retrievals are not as good as those obtained with the isothermal assumption. The $\chi^2/n$ is also higher for the single slab model than it is when the correct temperature profile is used.

The interpretation of transmission measurements without a good temperature constraint from eclipse is therefore something to be undertaken with caution.  Even if eclipse results are available and the dayside temperature is constrained, it is important to bear in mind that hot, short-period planets are often tidally locked, so we might expect some variation in temperature between the limb as seen in transmission and the star-facing disc as seen in eclipse; any assumptions made about heat redistribution will also affect the retrieval of gaseous abundances in transmission geometry. Some information on this score can be obtained using phase curves/transit mapping, as in the analyses of HD 189733b performed by \citet{maj12} and \citet{dewit12}, which provides information about the brightness temperature as a function of local time on the planet. The limitations of optimal estimation mean that it is not possible to use something like NEMESIS to independently constrain temperature and gaseous abundances from a primary transit spectrum, but the difference in $\chi^2/n$ between the different scenarios indicates that there is some information available in the spectrum about the temperature. This can be exploited by performing multiple retrievals with different temperature profiles and comparing the $\chi^2$, as suggested above. 

\begin{figure}
\centering
\includegraphics[width=0.4\textwidth]{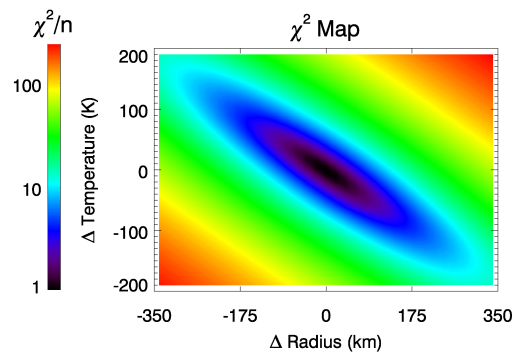}
\caption{A map of $\chi^2/n$ when the radius and temperature of the best-fit model to a noisy synthetic are perturbed from the correct value. There are clearly several possible solutions with $\chi^2/n$ of less than 2, indicating that there is degeneracy between planetary radius and temperature.\label{chisqmap}}
\end{figure}

We also find that there is degeneracy between the assumed planetary radius and the temperature in retrievals of transmission spectra. In Figure~\ref{chisqmap}, we show $\chi^2/n$ for a range of model fits where the planetary radius and temperature are perturbed from the known input value. It is clear that for a range of radii and temperatures, $\pm$150 km and $\pm$ 60 K respectively, a reasonable fit to the noisy synthetic is obtained. This degeneracy arises from the fact that increasing the planet radius and increasing the temperature have the same effect on a transmission spectrum, as both make the planet appear to be larger with an increased scale height. Therefore, a simpler way to account for a lack of information about temperature structure might be to adjust the modelled radius until $\chi^2/n$ is minimized; this would allow us to obtain accurate information about the atmospheric composition even if we do not know the temperature. 

\subsection{Warm Jupiter}
\begin{figure*}
\centering
\includegraphics[width=1.0\textwidth]{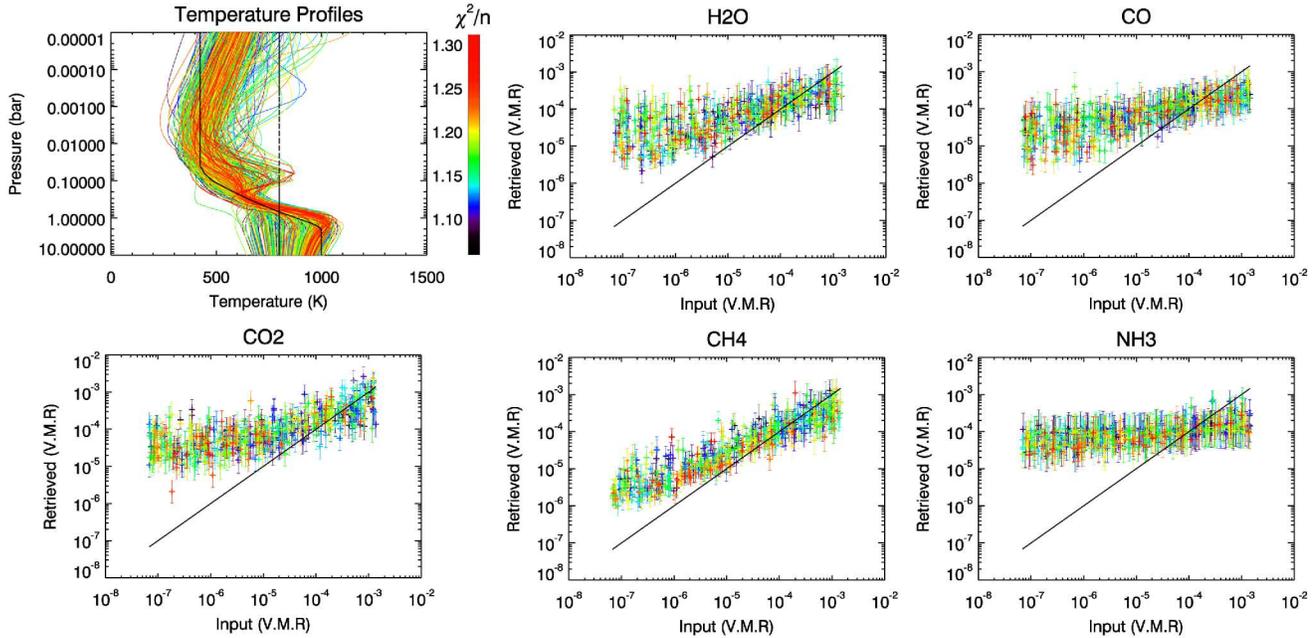}
\caption{Retrieval results for a warm Jupiter with a single eclipse; colours/linestyles as Figure~\ref{hotjupresults}.\label{warmjupresults}}
\end{figure*}

\begin{figure*}
\centering
\includegraphics[width=1.0\textwidth]{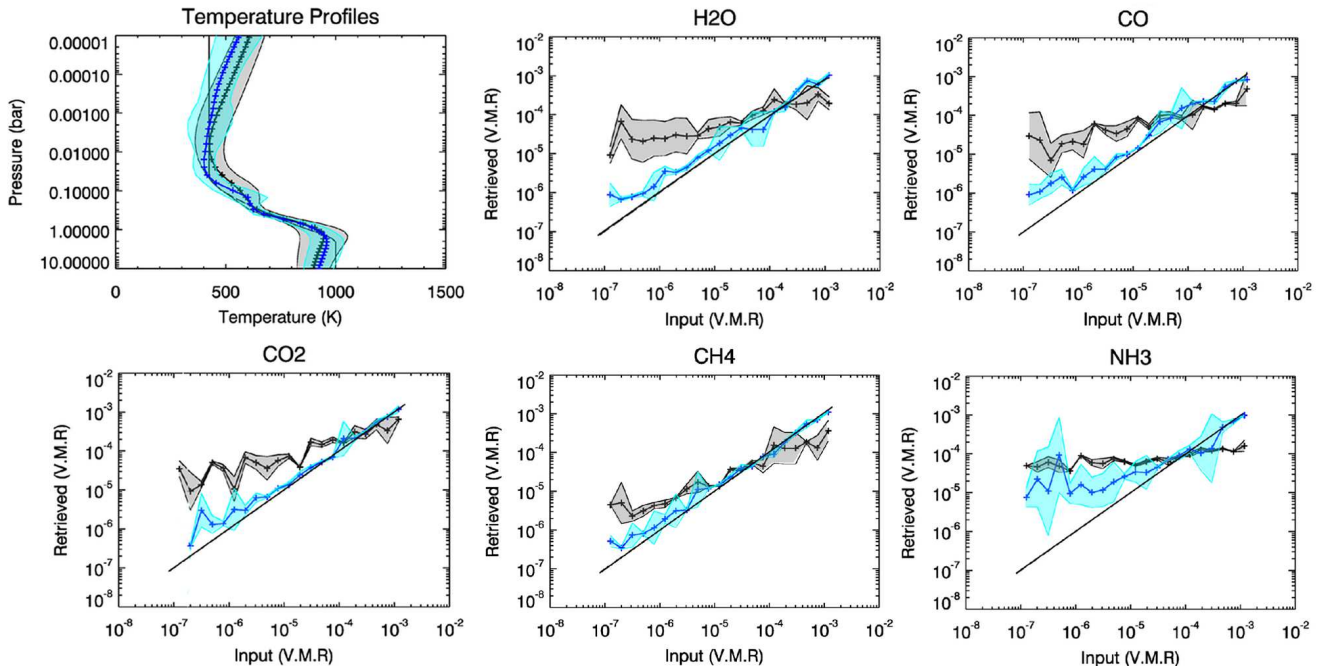}
\caption{Comparison of retrieval results for a warm Jupiter assuming errors for a single eclipse (black/grey) and 30 observations combined (turquoise/blue). Joined crosses are the mean over 100 retrievals, shaded areas are standard deviations. The solid black lines represent the input temperature and the 1:1 correspondence line in the temperature and gas plots respectively. The increase in SNR for 30 transits much improves the quality of the retrieval.\label{warmjup30x}}
\end{figure*}

\begin{figure*}
\centering
\includegraphics[width=0.8\textwidth]{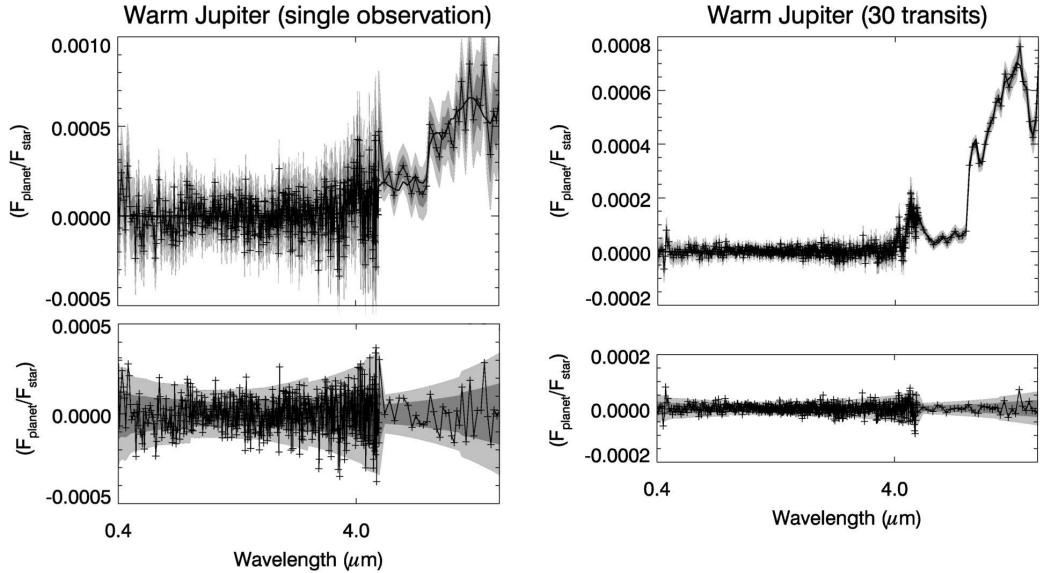}
\caption{Fits to noisy synthetic warm Jupiter eclipse spectra. The residuals are shown in the lower panels. The 1-$\sigma$ error bar is shown in dark gray shading, and the 2-$\sigma$ error in light gray. The joined crosses in the top panels show the noisy synthetic, and the solid black lines show the fitted spectra.\label{cold_j_examplespec}}
\end{figure*}

\begin{figure*}
\centering
\includegraphics[width=1.0\textwidth]{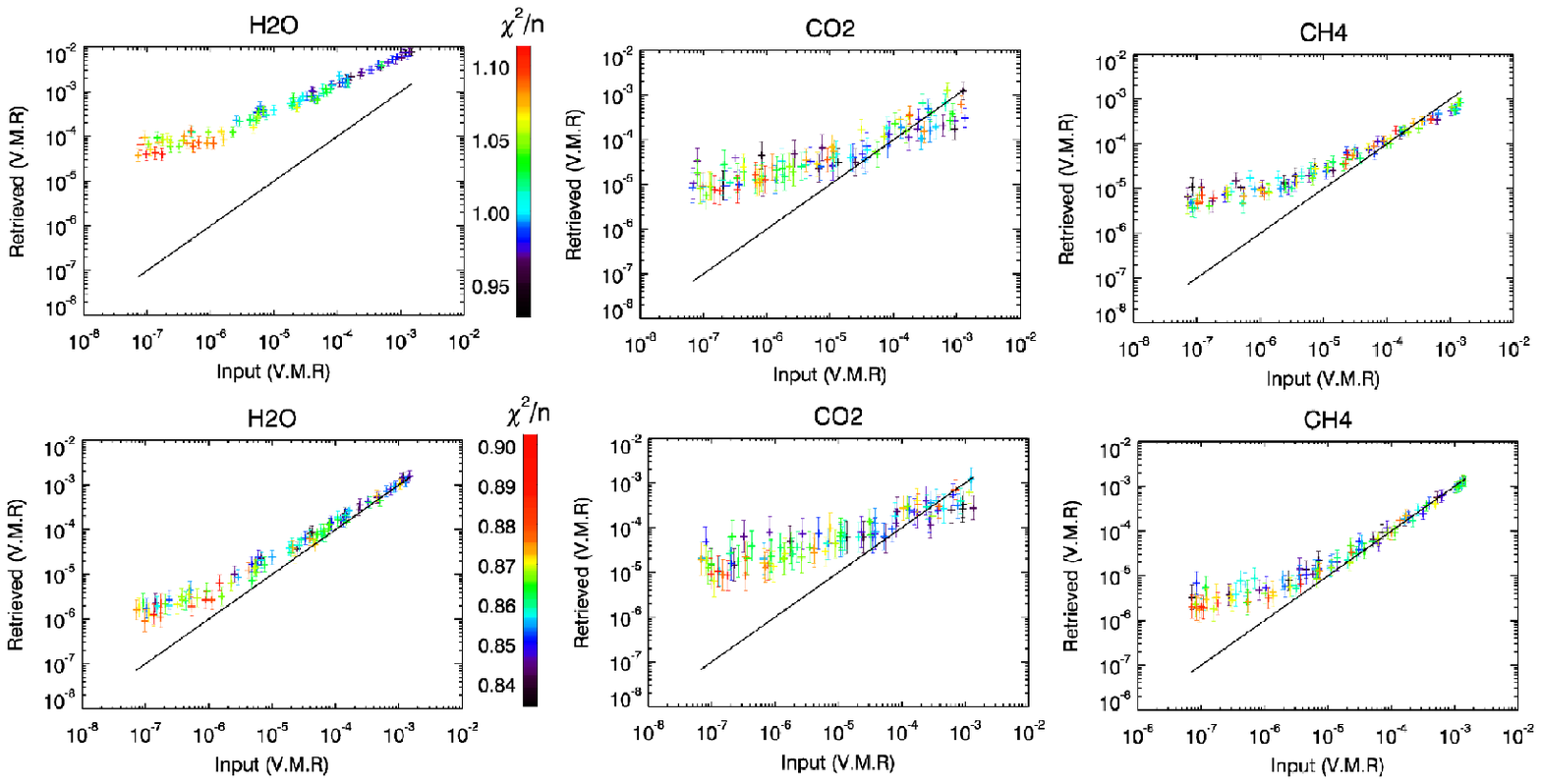}
\caption{Retrievals of H$_2$O, CO$_2$ and CH$_4$ from single transmission spectra for a warm Jupiter with no temperature inversion. In the upper panels, a constant temperature of 600 K is assumed; in the lower panels, the input temperature profile is assumed. Colours are as Figure~\ref{hotjupresults}.\label{coldjupgasprimary}}
\end{figure*}

\begin{figure*}
\centering
\includegraphics[width=1.0\textwidth]{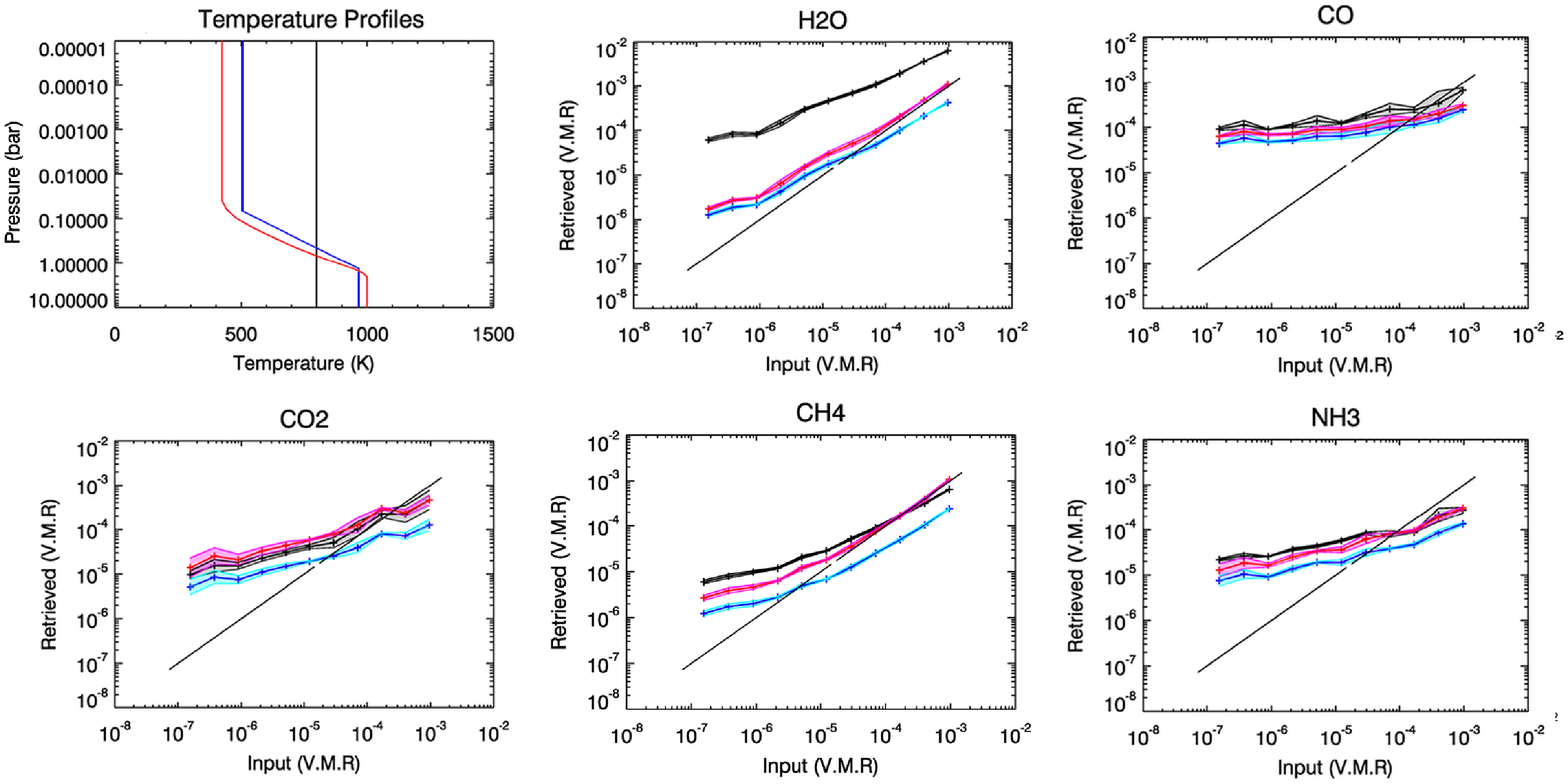}
\caption{As Figure~\ref{hotjadiabattest} for the warm Jupiter case.\label{warmjadiabattest}}
\end{figure*}

We now consider the case of a somewhat cooler Jupiter-size planet. The temperature structure is similar to that of the hot Jupiter without a temperature inversion, but the temperature at 1 bar is 1000 K instead of 2500 K. This reduces the planet/star contrast and the signal to noise in eclipse, and in Figure~\ref{warmjupresults} we see that the retrieval for a single eclipse is poor. 

The signal to noise could be improved in this case if several observations were added together. We repeat the test with the added noise reduced by a factor of $\sqrt{30}$, which is the equivalent of observing 30 transits and combining the results assuming that all error is random rather than systematic; as expected, the quality of the retrieval is significantly improved (Figure~\ref{warmjup30x}). The spectra for both cases are shown in Figure~\ref{cold_j_examplespec}. Based on the input temperature profile this planet would have an equilibrium temperature of around 600 K, which corresponds to an orbital radius of 0.3 AU for a sun-like star; it would therefore take around 60 days to complete one orbit. If at least 30 transits are required before sufficient signal to noise is achieved, then such a planet would require observation over EChO's proposed five-year lifetime. This means that retrievals of similar planetary atmospheres are a possibility, but the periods of planets like this would need to be well characterised prior to the start of the mission, perhaps using ground-based telescopes, to ensure that the efficiency of observation is maximised. It is worth noting that varying stellar activity may, as mentioned by \citet{tinetti10}, complicate the process of averaging over several observations. 

\begin{figure}
\centering
\includegraphics[width=0.4\textwidth]{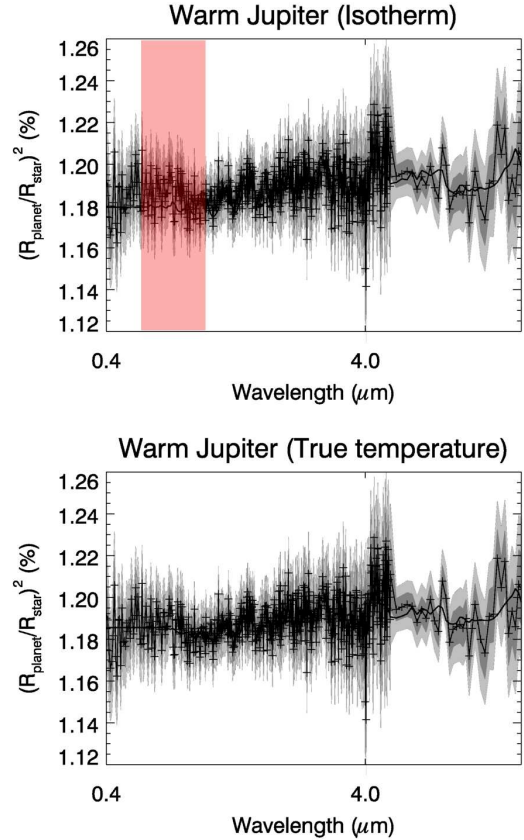}
\caption{Examples of fits to a noisy synthetic warm Jupiter transmission spectrum. (as Figure~\ref{hot_j_examplespec_pt}).\label{warm_j_examplespec_pt}}
\end{figure}

The quality of transmission observations is also reduced for a cooler planet (Figures~\ref{coldjupgasprimary} and~\ref{warmjadiabattest}); the magnitude of the transmission signal depends on the scale height of the atmosphere, which is proportional to temperature, so the warm Jupiter has a compressed atmosphere relative to the hot Jupiter. Example spectra for two different temperature assumptions are shown in Figure~\ref{warm_j_examplespec_pt}, along with those for the hot Neptune case. As with the hot Jupiter, a good prior constraint on the temperature is necessary for a reasonable retrieval of gaseous abundances, so it will probably be necessary to perform multiple retrievals with different temperature profiles and search from the minimum $\chi^2$, as suggested for the hot Jupiter. A good temperature constraint is harder to obtain from eclipse measurements for a cooler planet, and the difference in $\chi^2/n$ between good and poor solutions is smaller for spectra with larger error bars, so in both transit geometries cooler planets present an observational challenge.

\subsection{Hot Neptune}

\begin{figure}
\centering
\includegraphics[width=0.4\textwidth]{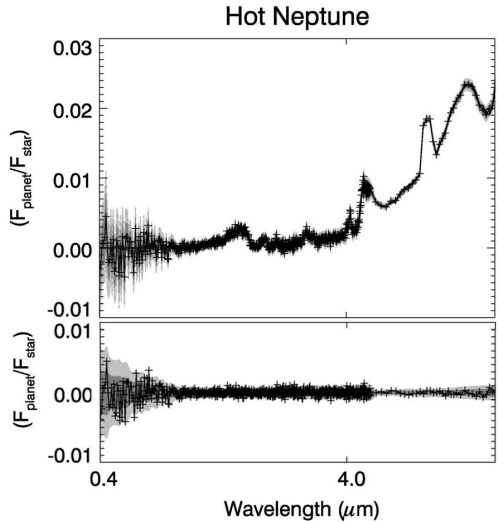}
\caption{Fits to a noisy synthetic hot Neptune eclipse spectrum, as Figure~\ref{cold_j_examplespec}.\label{hot_n_examplespec}}
\end{figure}

\begin{figure}
\centering
\includegraphics[width=0.4\textwidth]{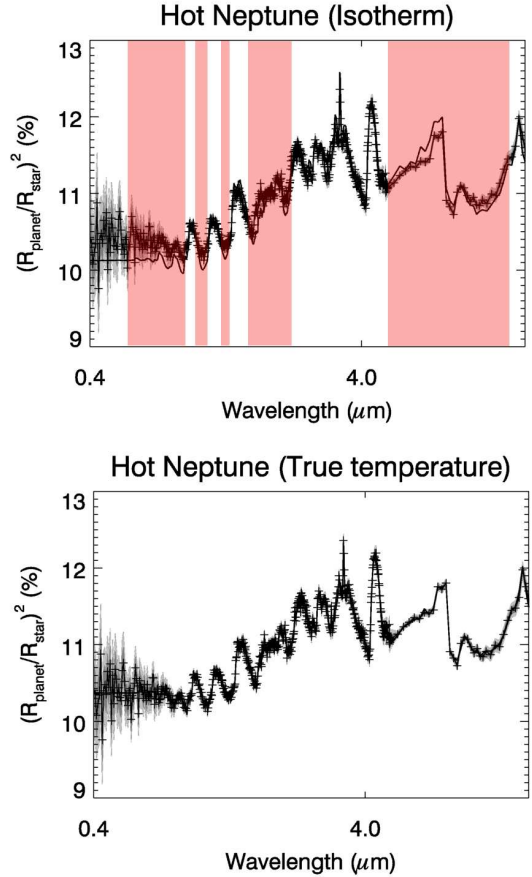}
\caption{Examples of fits to a noisy synthetic hot Neptune transmission spectrum. (as Figure~\ref{hot_j_examplespec_pt}).\label{hot_n_examplespec_pt}}
\end{figure}

\begin{figure*}
\centering
\includegraphics[width=1.0\textwidth]{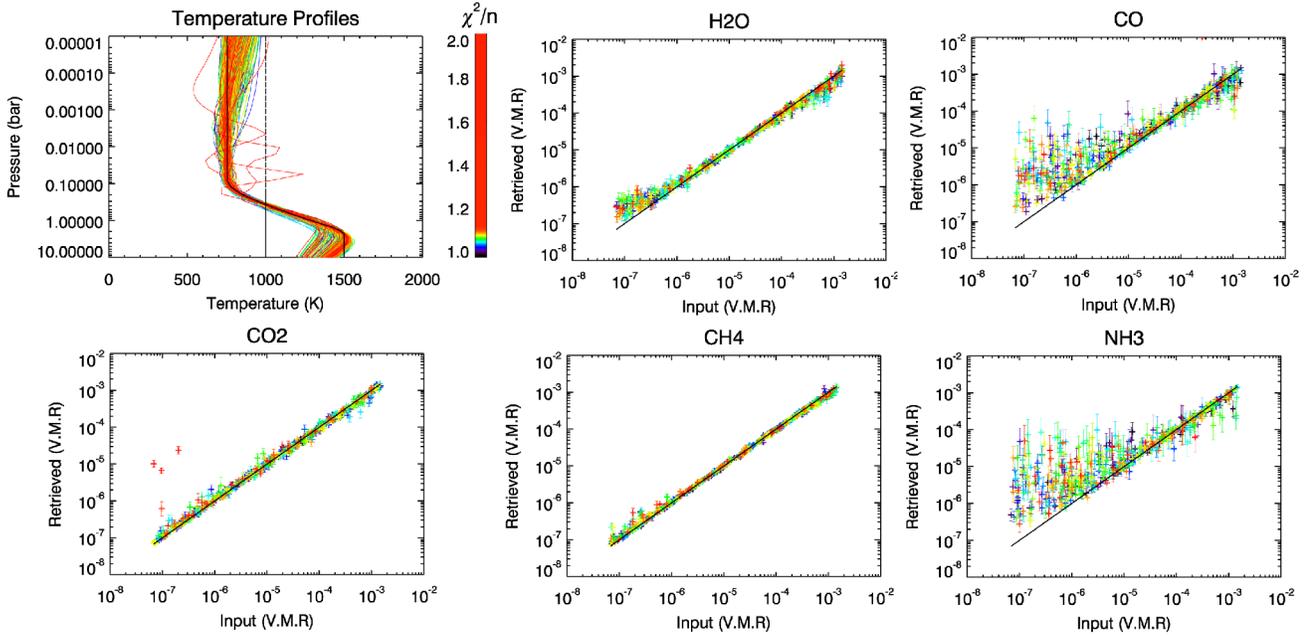}
\caption{Eclipse retrieval results for a hot Neptune orbiting an M star; colours/linestyles as Figure~\ref{hotjupresults}.\label{hotnepresults}}
\end{figure*}

Finally, we consider a planet approximately the size of Neptune orbiting an M dwarf. Even though this planet is smaller and cooler (1500 K in the deep atmosphere) than the hot Jupiter we consider, because it is orbiting a smaller, cooler star the contrast ratios are better (Figure~\ref{hot_n_examplespec}). We can successfully retrieve the temperature profile and VMRs of H$_2$O, CO$_2$ and CH$_4$ assuming noise levels for a single transit (Figure~\ref{hotnepresults}). We can also retrieve VMRs of CO greater than 10 ppmv. 

\begin{figure*}
\centering
\includegraphics[width=1.0\textwidth]{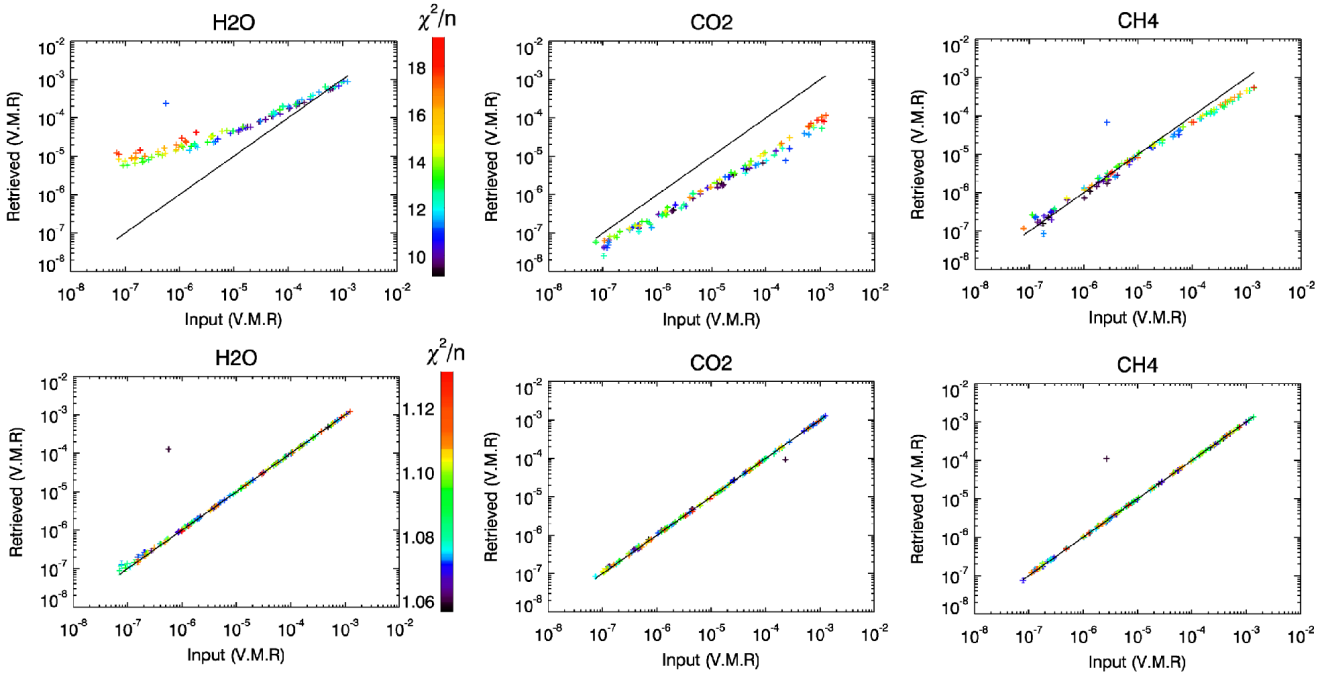}
\caption{Transmission retrieval results for a hot Neptune orbiting an M dwarf (as Figure~\ref{hotjupgasprimary}).\label{hotnepgasprimary}}
\end{figure*}

The hot Neptune has an extended H$_2$-He atmosphere and is orbiting a
relatively small star, so the SNR in transmission is very high
(Figure~\ref{hot_n_examplespec_pt}). As for the hot Jupiter cases, when
the assumed temperature profile is correct the retrievals of H$_2$O,
CO$_2$ and CH$_4$ VMRs are extremely accurate, but
when an isothermal temperature profile is used the retrieved value is
incorrect and the fit is poor (Figures~\ref{hotnepgasprimary} and~\ref{hot_n_examplespec_pt}
). As with the hot Jupiter, the small errors
on the spectrum mean that poor assumptions about the temperature
profile are associated with high $\chi^2$, so it should be possible to
achieve a good retrieval of gas properties by varying the assumed
temperature and repeating the retrieval.

\section{Discussion}

For the simple model planets described above, our results indicate that it will be possible for the EChO space telescope, and any other telescopes with a similar spectral range and an SNR/spectral resolution as good as EChO's, to reliably retrieve temperature as a function of pressure and VMRs for H$_2$O, CO$_2$ and CH$_4$, with upper limits for CO and NH$_3$. We now consider the validity of some of the simple assumptions made here, and discuss how changes in these might affect our conclusions.

\subsection{Planetary mass and radius}

We have assumed that we have perfect knowledge of the planetary mass and radius. The methods used to calculate these parameters for transiting exoplanets mean that their precision depends on the precision to which we know the mass and radius of the parent star. Errors quoted in \citet{demory09} for Proxima Centauri (GJ 551) are approximately 5 \%, therefore we may expect at least a 5 \% error in measurements of planetary mass and radius. We also assume the measured transit radius is equivalent to the radius at the bottom of our atmosphere, i.e. at 10 bar pressure, which may not be the case in reality as the atmosphere is likely to be opaque to higher altitudes in some wavelengths. The absolute radius derived from the transit depth averaged over all visible wavelengths of our model hot Jupiter is 2\% greater than the specified radius at 10 bar. 

The worst case scenario is that the derived mass is too large/small when the derived radius is too small/large, as this produces the largest error in the calculated gravitational acceleration. The scale height and dry adiabatic lapse rate, and therefore the temperature profile, depend on the gravitational acceleration, so large variations in $g$ can make a significant difference to spectra. We test the effect of this by comparing retrievals assuming 1) the actual input radius and mass, 2) 95\% of the mass and 105\% of the radius and 3) 105\% of the mass and 95\% of the radius. The retrieved temperature profiles in eclipse are compared below, along with the $\chi^2/n$ values.

\begin{figure*}
\centering
\includegraphics[width=1.0\textwidth]{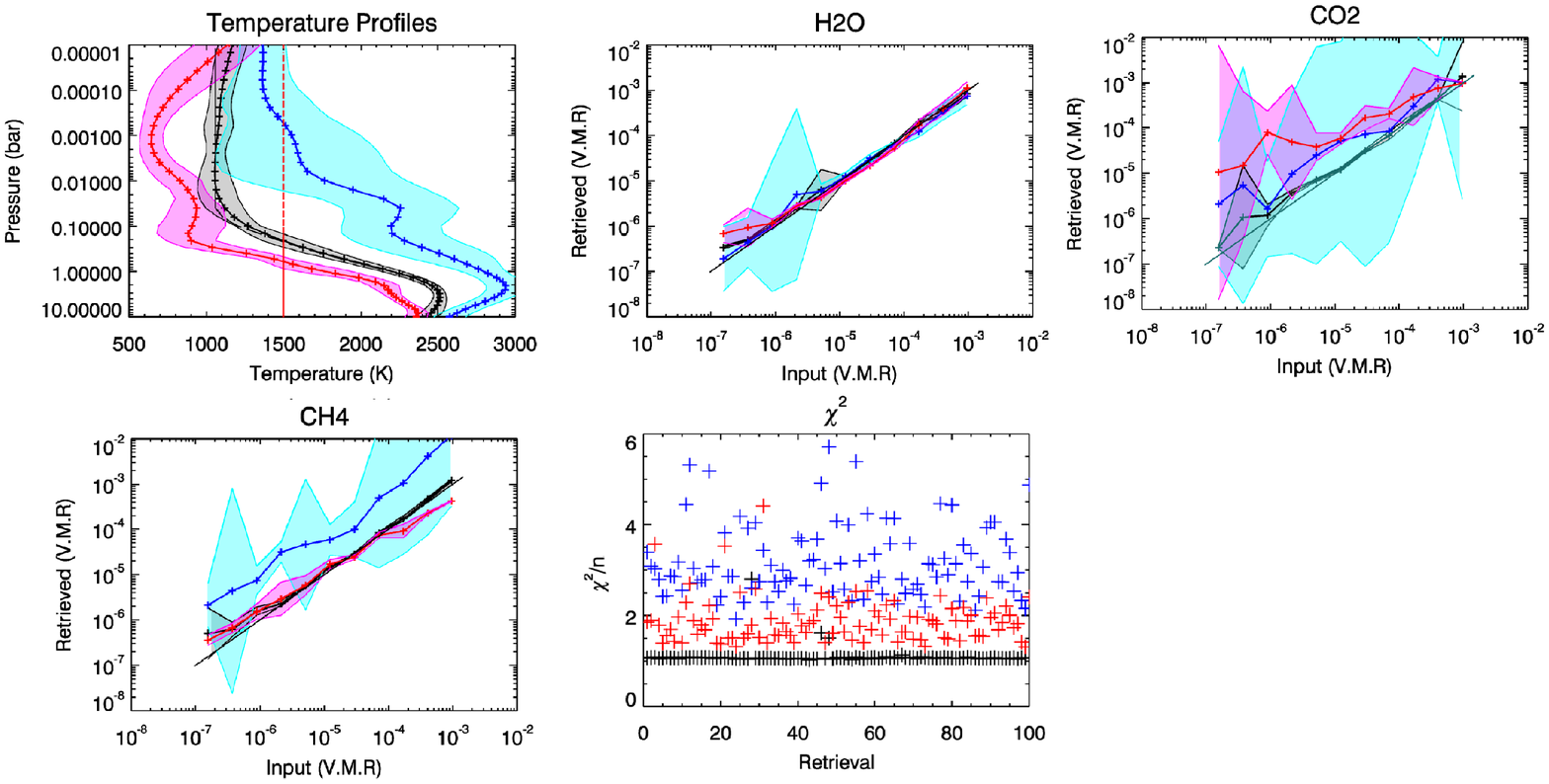}
\caption{A comparison between retrievals with the correct planet mass and radius (black/grey) and retrievals with a low mass/high radius (blue/turquoise) and a high mass/low radius (red/fuschia). The joined crosses are the mean values and shaded areas represent the standard deviation over 100 retrievals. The $\chi^2/n$ is also shown.\label{juphot_grav_disc}}
\end{figure*}

These extreme cases of inaccuracy in the planetary mass and radius make a large difference to the retrieved temperature, CO$_2$ and CH$_4$ values. A too-low gravity has a more significant effect that a too-high gravity. Using the wrong values for mass and radius increases the $\chi^2/n$ parameter to a value above 1, which indicates that the fit to the noisy synthetic spectra becomes poor; the retrieved temperature profiles are also less smooth, particularly for the case where the gravity is low. This means that it should be possible to reject a result where the wrong mass and radius are assumed for a real planet, and indeed it may be possible to constrain the correct range of $g$ by minimising $\chi^2/n$ over different values of mass and radius.

We also test the impact of errors in the mass and radius on results in transmission geometry. The same deviations as with the eclipse case mean that NEMESIS is unable to produce a fit to the spectrum at all, resulting in $\chi^2/n$ values of several thousands. As with previous results of a similar nature, the extremely high values of $\chi^2/n$ indicate that such a retrieval would not be accepted, and so even if errors on the mass and radius of the planet are present it should still be possible to obtain sensible constraints through trial and error of different planetary properties. However, as shown in Figure~\ref{chisqmap} there is significant degeneracy between radius and temperature in primary transit, and we would expect degeneracy between $g$ and temperature too as both affect the atmospheric scale height ($H=kT/mg$), so constraining $g$ in this way without any constraint on temperature would be difficult for primary transits. 

Even if errors in the planetary mass and radius are as great as 5\% when these are derived from stellar mass and radius, it is possible to place more precise constraints on the planetary surface gravity by calculating it directly from transit and radial velocity observables \citep{south07}. A large source of error in planetary mass and radius is a lack of constraint on stellar mass and radius, whereas this alternative method does not require knowledge of these parameters. The large effect of uncertainty in $g$ on the retrievals discussed here means that the method of Southworth et al. is likely to be very important for calculating the gravitational acceleration of new planet candidates.

\subsection{Systematic noise}

\begin{figure}
\centering
\includegraphics[width=0.4\textwidth]{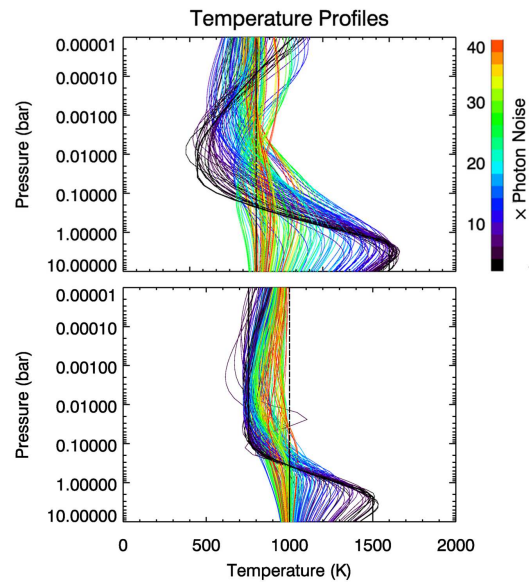}
\caption{Effect of increasing the eclipse spectrum noise on the temperature retrieval for a hot Jupiter with a temperature inversion, orbiting a G2 star (upper panels) and a hot Neptune orbiting an M5 star (lower panels). Colours correspond to the noise level relative to the photon-limited case.\label{rednoise_temp_eclipse}}
\end{figure}

\begin{figure*}
\centering
\includegraphics[width=1.0\textwidth]{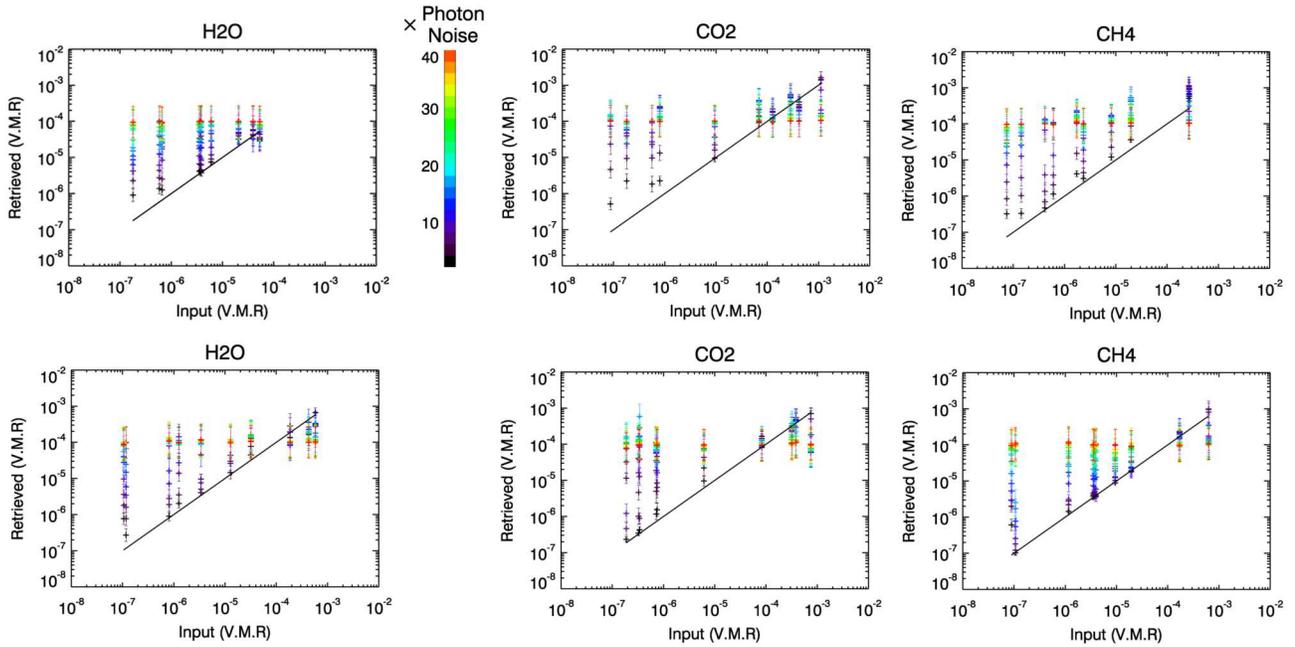}
\caption{Effect of increasing the eclipse spectrum noise on the gas abundance retrieval for a hot Jupiter with a temperature inversion, orbiting a G2 star (upper panels) and a hot Neptune orbiting an M5 star (lower panels). \label{rednoise_eclipse}}
\end{figure*}

\begin{figure*}
\centering
\includegraphics[width=1.0\textwidth]{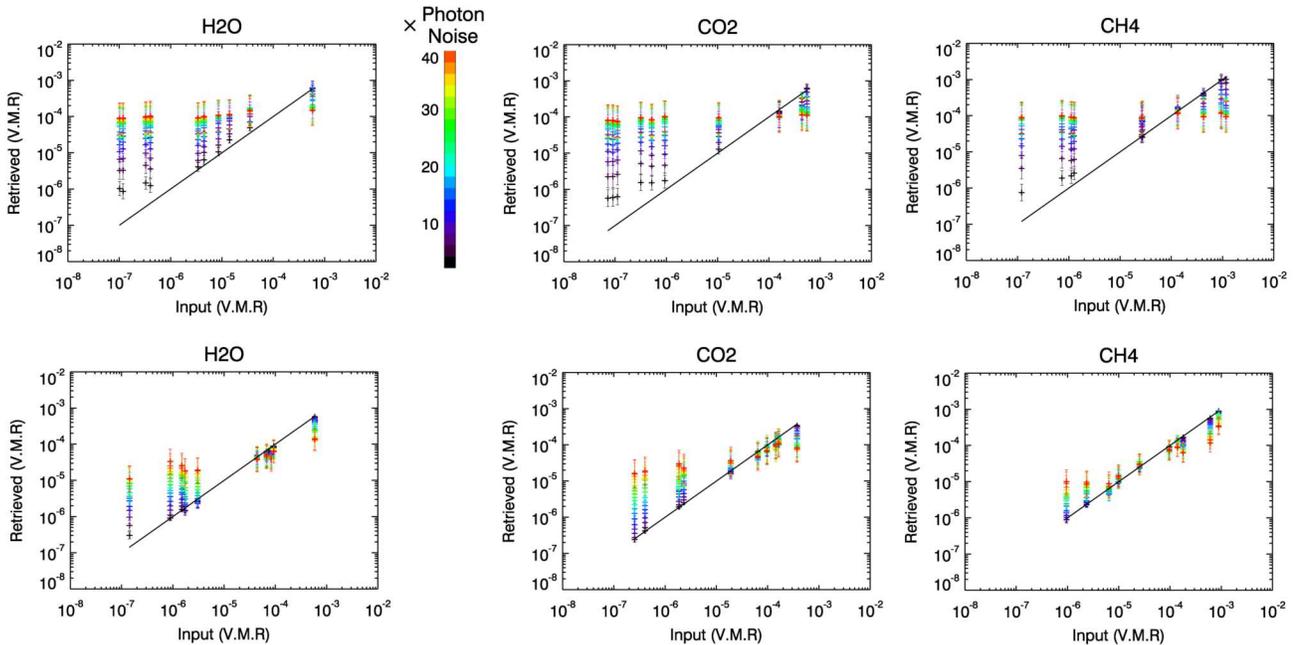}
\caption{As Figure~\ref{rednoise_eclipse}, but for transmission.\label{rednoise_transit}}
\end{figure*}

As mentioned in Section~\ref{noisecalc}, the analysis so far has assumed photon-limited noise, which is a perfect case; in reality, we expect some systematic contribution to the noise from the instrument and from sources such as stellar variability. It is a useful exercise to investigate the effect of multiplying the photon noise by a range of factors, and determine at what point the signal is lost for each of the cases above. The results of these tests are shown in Figures~\ref{rednoise_temp_eclipse} --- ~\ref{rednoise_transit} for the hot Jupiter and hot Neptune cases, where the colours correspond to the increase in the noise level. Noise levels up to 40$\times$ the photon noise were tested in order to investigate the point at which the retrieval completely fails, but more likely values are in the 2---4$\times$ range. It can be seen that in eclipse, a factor 4 increase (very dark purple) from the photon noise limit does not substantially affect the quality of the temperature retrieval, and a factor 2 increase (black) does not affect the gas volume mixing ratio retrieval for higher abundances. In transmission, the effect for a factor 2 increase is not significant but a factor 4 increase does start to affect the retrieval accuracy. If however a factor 3 increase represents a worst-case scenario (see Section~\ref{noisecalc}), this indicates that the inclusion of a reasonable level of systematic noise does not substantially change the conclusions we arrive at. For a factor 2 increase in the noise level for the hot Jupiter, we can only detect H$_2$O down to an abundance of 1 ppmv, CO$_2$ down to 5 ppmv and CH$_4$ down to 0.5/5 ppmv (eclipse/transmission); for the hot Neptune, the detection limit for a factor 2 increase is unchanged, but for a factor 4 it increases to 5 ppmv in transmission and eclipse for H$_2$O and CO$_2$, and 5 ppmv for CH$_4$ in eclipse. At around 30$\times$ the photon noise there is not enough signal remaining for the retrieval to move away from the \textit{a priori}.

We have not shown the effect of increasing the noise on the warm Jupiter spectra because we are already considering adding together multiple transits to gain sufficient signal to noise; the effect of systematic noise would simply be to increase the number of transits required to achieve a given retrieval accuracy, although in this case a small noise increase may result in the mission lifetime being insufficient for the number of observations required. However, the averaging process introduces another source of systematic noise, that of changing stellar activity. The majority of stars have spots on their surfaces, which tend to be cooler than the rest of the disc and also have different spectral signatures. The fractional spot coverage is unlikely to change on the timescale of a single transit, but between visits of around 60 days apart, as for the warm Jupiter, we may expect it to vary. Star spots can introduce error into planetary transmission spectra because the part of the stellar disc occulted by the planet may not be representative of the whole. \citet{pont12} discuss the effects of unocculted star spots on transmission spectra, and show that the magnitude of the effect is most significant at optical wavelengths. 

\begin{figure}
\centering
\includegraphics[width=0.4\textwidth]{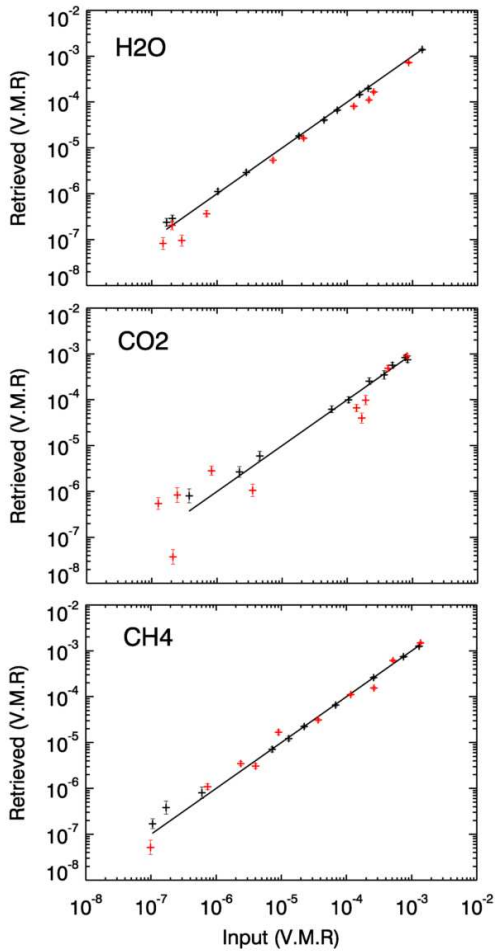}
\caption{Warm Jupiter transmission results over 30 transits where the star is unspotted (black points) and where spot-like effects have been added to the synthetic spectrum (red points). Both retrievals have assumed that the star is unspotted. It can be seen that star spots are an extra source of error, but retrievals are still possible.\label{starspots}}
\end{figure}

Assuming that the ratio of the spot spectrum to the unspotted solar spectrum can be approximated as a ratio of Planck functions at 4800 and 5800 K respectively, we use the formulae given in \citet{pont12} to calculate the adjusted planet/star radius ratio for a range of spot flux dimmings at 0.6 $\upmu$m. We assume, based on the magnitudes given in \citet{pont12}, that the 1$\sigma$ flux dimming is 0.25 \% at 0.6 $\upmu$m, and we generate 30 spectra with different amounts of star spot dimming, plus photon noise, for each test atmospheric composition. We then average over these 30 spectra and perform a retrieval on ten examples, enabling us to quantify the increase in error due to stellar variability. The results of this test are shown in Figure~\ref{starspots}. It can be seen that the error on the retrieved H$_2$O and CH$_4$ VMRs increases by around a factor 3, whereas the error on the CO$_2$ VMR could be as much as an order of magnitude. However, no attempt has been made to account for the spot effect in the retrieval, so this could conceivably be improved upon if the stellar flux is monitored and the inferred spot coverage corrected for, as described in \citet{pont12}.

If only one or two hot Neptunes around M dwarfs are observed by EChO, it is likely that these would also be observed several times. We perform a similar test for the sensitivity to star spots in this case, with a spot temperature of 2000 K compared with a stellar temperature of 3000 K, and find that the effect is negligible due to the much larger signal in transmission than that of the warm Jupiter.

We do not consider occulted spots in this analysis because these should have observable signatures in the visible wavelength lightcurves \citep{pont12}, so are somewhat easier to correct for. We also do not consider the effect of star spots on eclipse measurements because we do not expect significant spot evolution during an eclipse, and as the full stellar disc is visible throughout the spot distribution does not affect the measurement.

\subsection{Missing absorbers}

\begin{figure}
\centering
\includegraphics[width=0.4\textwidth]{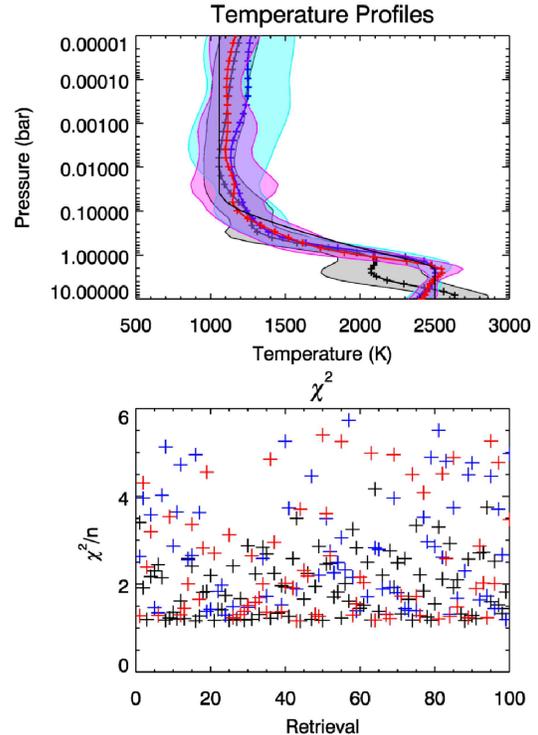}
\caption{Variation in retrieved temperature if H$_2$O (black/grey), CO$_2$ (blue/turquoise) and CH$_4$ (red/fuschia) are not included in the retrieval model. The joined crosses are the mean values and shaded areas represent the standard deviation over 100 retrievals. The $\chi^2/n$ is also shown for each case.\label{temp_ignoregas}}
\end{figure}

We test the effect of removing gaseous absorbers from the retrieval model that were included in the synthetic spectrum calculation. We perform this test for H$_2$O, CO$_2$ and CH$_4$ as these are the gases that produce the largest effect on the spectra, as evidenced by the reliability of the retrievals for these gases. If any one of these gases is ignored, the temperature retrieval becomes less accurate, with H$_2$O having the most significant effect. The $\chi^2/n$ value is increased from the values when all the gases are included, but the correct shape of the temperature profile is still retrieved. As gas absorption line data at high temperatures is often incomplete, it is necessary to be aware that missing lines can impact the reliability of temperature retrieval.

\subsection{Altitude variation of absorbers}
\label{altvar}

\begin{figure}
\centering
\includegraphics[width=0.4\textwidth]{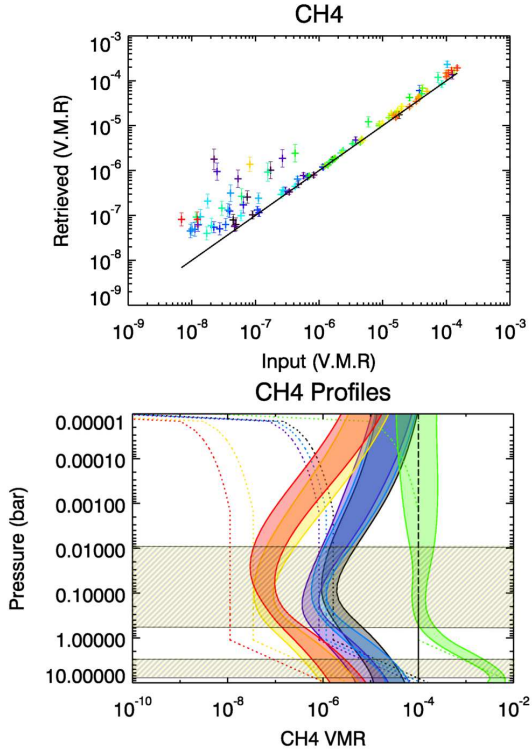}
\caption{Top: 100 CH$_4$ retrievals assuming constant mixing ratio. Colours correspond to $\chi^2/n$ as previously. The input values are the VMRs at 0.1 bar. Bottom: 6 examples of a continuous CH$_4$ retrieval. The enclosed shaded areas represent the error envelope of each retrieval. The dotted profiles of the same colours are the corresponding model inputs. The black dashed line is the \textit{a priori} profile. The hatched areas indicate the atmospheric pressure regions for which the CH$_4$ VMR can be retrieved in the majority of cases.\label{contch4}}
\end{figure}

We also test the effect of including more realistic profiles of gas VMR as a function of altitude. \citet{moses11} use a 1D photo- and thermo-chemical model to calculate the expected abundances of trace absorbers on HD 189733b and HD 209458bb. They find that CH$_4$ and NH$_3$ are expected to be photochemically removed at high altitudes. We test the effect on our results of including a more realistic vertical CH$_4$ profile, based on the prediction of \citet{moses11}.

\begin{figure}
\centering
\includegraphics[width=0.4\textwidth]{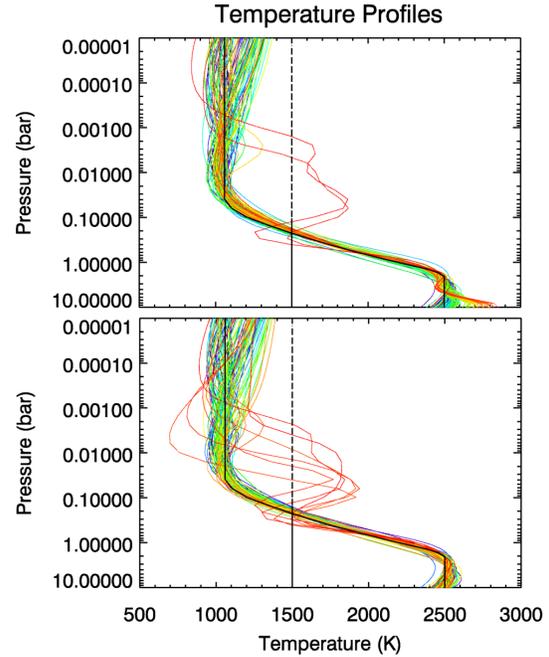}
\caption{100 temperature retrievals when a more realistic CH$_4$
  vertical profile is included in the model atmosphere. In the upper
  panel is the temperature retrieved when a constant CH$_4$ profile is
  still assumed in the retrieval; in the lower is the temperature retrieved when CH$_4$ VMR is retrieved as a function of pressure.\label{temp_contch4}}
\end{figure}

\begin{figure}
\centering
\includegraphics[width=0.4\textwidth]{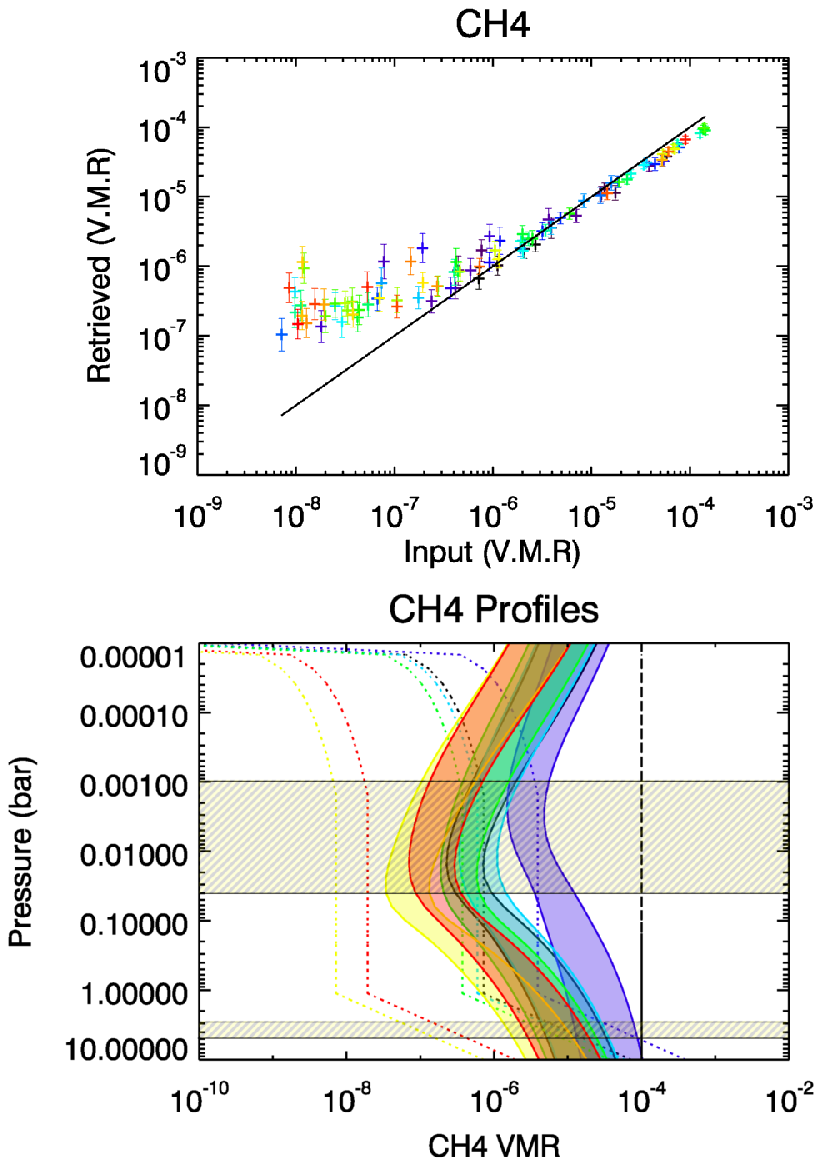}
\caption{As Figure~\ref{contch4} but for transmission.\label{contch4pt}}
\end{figure}

\begin{table*}
\begin{minipage}{126mm}
\begin{tabular}[c]{|c|c|c|c|}
\hline
Quantity & Hot Jupiter & Warm Jupiter & Hot Neptune\\
\hline
Temperature & 200 K (1 bar---1 mbar) & 250 K (0.5 bar---5 mbar) & 100 K (1 bar---1 mbar) \\
H$_2$O & 2$\times$ ($>$0.5 ppmv) & 4$\times$ ($>$50 ppmv) & 2$\times$ ($>$0.5 ppmv) \\
CO$_2$ & 3$\times$ ($>$10 ppmv), 6$\times$ ($>$0.1 ppmv) & 5$\times$ ($>$50 ppmv) & 2$\times$ ($>$0.1 ppmv)\\
CO & 5$\times$ ($>$100 ppmv) & Not detectable & 3$\times$ ($>$10 ppmv) \\
CH$_4$ & 2$\times$ ($>$1 ppmv), 3$\times$ ($>$0.1 ppmv) & 4$\times$ ($>$50 ppmv) & 1.5$\times$ ($>$0.1 ppmv)\\
NH$_3$ & 5$\times$ ($>$100 ppmv) & Not detectable  & 3$\times$ ($>$100 ppmv) \\
\hline
\end{tabular}
\caption{Information available from a single observation for each of the test planets in eclipse. These precisions are only valid for certain pressures (for temperature) and concentrations (for gases), which ranges are given in parentheses for each case. Precisions are quoted as $\pm$K for temperature. As the gas abundances vary over several orders of magnitude, the precisions are given as multiples of the abundance; for example, the H$_2$O VMR can be retrieved to within a factor 2 of the true value for a hot Jupiter, provided there is more than 0.5 ppmv present. The values quoted take into account the errors due to retrieval degeneracy, which is the dominant source of error, by calculating the precision based on the 2$\sigma$ deviation from the true value over all retrievals.\label{results_occultation}}
\end{minipage}
\end{table*}

\begin{table*}
\begin{minipage}{126mm}
\begin{tabular}[c]{|c|c|c|c|}
\hline
Quantity & Hot Jupiter & Warm Jupiter & Hot Neptune\\
\hline
H$_2$O & 1.5$\times$ ($>$0.5 ppmv) & 1.5$\times$ ($>$10 ppmv) & 1.5$\times$ ($>$0.1 ppmv) \\
CO$_2$ & 1.5$\times$ ($>$0.5 ppmv) & 2$\times$ ($>$50 ppmv) & 1.5$\times$ ($>$0.1 ppmv)\\
CO & Not detectable & Not detectable & 1.5$\times$ ($>$10 ppmv) \\
CH$_4$ & 1.5$\times$ ($>$0.5 ppmv) & 2$\times$ ($>$10 ppmv) & 1.5$\times$ ($>$0.1 ppmv)\\
NH$_3$ & 3$\times$ ($>$100 ppmv) & Not detectable  & 1.5$\times$ ($>$5 ppmv) \\
\hline
\end{tabular}
\caption{Information available from a single observation for each of the test planets in transmission. Perfect knowledge of the temperature profile has been assumed. Otherwise everything is as Table~\ref{results_occultation}.\label{results_transmission}}
\end{minipage}
\end{table*}

We first test the effect of including a more complex CH$_4$ profile in the model but performing the retrieval in the same way as before, i.e. only retrieving a single multiplying factor on a constant VMR. Despite this crude assumption, we can accurately retrieve the abundance of CH$_4$ at 0.1 bar (Figure~\ref{contch4}, upper panel), which is within the altitude range of maximum sensitivity. Ignoring the detail of the CH$_4$ vertical profile has very little effect on the retrieval of other quantities, except for increased error on the temperature close to the 10 bar level, which is reassuring as it implies little degeneracy between the sort of altitude variation in gas abundance that we would expect and changes in temperature structure (Figure~\ref{temp_contch4}, upper panel). 

We also investigate the possibility of obtaining information about the vertical profile of CH$_4$ from eclipse measurements by performing a continuous retrieval of CH$_4$ VMR as a function of pressure. Whilst this removes the error in retrieved temperature in the deep atmosphere (Figure~\ref{temp_contch4}, lower panel), it is apparent that not much information about the CH$_4$ profile is available from the spectrum (Figure~\ref{contch4}, lower panel). The continuous retrieval only matches the input at two pressure levels, and everywhere else relaxes back to the \textit{a priori} solution, so we do not obtain much more useful information by increasing the complexity of the retrieval even if absorber abundances are variable with altitude. $\chi^2/n$ is also slightly higher on average (1.1 rather than 1.0) for the continuous retrieval versus the constant VMR case because retrieving a continuous profile of CH$_4$ reduces the number of degrees of freedom; a continuous retrieval has more independently varying parameters than a multiplying factor. This indicates that the improvement in the fit to the spectrum, if any, is not sufficient to justify the increased number of retrieved parameters. Therefore, the simpler approach of assuming a vertically constant VMR for all gases is preferred for eclipse retrievals.

We perform the same test for the transmission case (Figure~\ref{contch4pt}). The results are similar to those for eclipse, with a limited amount of vertical information available, but the altitudes of sensitivity are slightly higher in transmission compared with eclipse; a combination of transmission and eclipse measurements could provide enough information to constrain variation with altitude, provided the temperature variation between the dayside and terminator regions is not significant. In both cases, the retrieval ceases to be accurate for tropospheric VMRs lower than 10$^{-6}$, so for values below this an upper limit only would be achievable.

\section{Conclusions}

We find that for a hot Jupiter orbiting a sun-like star and a hot Neptune orbiting a small M-dwarf a single eclipse observation by an EChO-like telescope is sufficient to provide good constraints on temperature structure and volume mixing ratios of H$_2$O, CO$_2$ and CH$_4$. Some constraint is also possible for CO and NH$_3$ VMRs, although when these values are low an upper limit only is achievable. The lower sensitivity to temperature structure in transmission observations means that independently constraining the temperature and gas VMRs from transmission spectra is not possible, but nonetheless errors in the assumed temperature structure do have a significant effect on the retrieved gas VMRs. The constraints on the gas VMRs are good provided the temperature structure has already been accurately retrieved from eclipse observations, so transmission and eclipse spectra combined could provide information about potential changes in atmospheric composition between the dayside and terminator. This approach of course relies on the assumption that the temperature profile at the terminator is the same as, or at least similar to, the dayside averaged temperature profile that is retrieved from the eclipse spectra; whether or not this is the case depends heavily on the efficiency of heat transport from the day to the night side of the planet, which is not known \textit{a priori}. A combination of transit mapping \citep{dewit12}, secondary eclipse and transmission measurements from a single orbit would therefore provide a wealth of information about hot Jupiters and Neptunes. These findings are summarised in Tables~\ref{results_occultation} and~\ref{results_transmission}.

We find that there is little dependence on the chosen \textit{a priori} temperature profile for eclipse retrievals, provided there is information contained in the spectrum; this demonstrates the reliability of the retrieval method used. We also find that retrievals are not very sensitive to vertical variation in the abundance of CH$_4$, a gas expected to vary as a function of pressure on planets like HD 189733b \citep{moses11}, justifying the common assumptions of constant VMRs for trace absorbers. The omission from the retrieval model of absorbers that are present in the original model atmospheres does however reduce the accuracy of the retrieved temperature profile, suggesting the need for caution in interpreting results where absorption line data is known to be incomplete. Theoretical programmes such as the Exomol project \citep{tenn12} and new experimental results (e.g. \citealt{har11}) can increase the robustness and reliability of retrieval results by improving the available line data.

Errors in the measured radius and mass of exoplanets could have an adverse effect on retrieval quality. A 5\% error in the mass and a 5\% error in the opposite sense in the radius result in the retrieved temperature profile in eclipse being wrong by as much as 500 K, which in turn impacts the retrieval of trace gas abundances. Similar errors in mass and radius make it impossible to produce a reasonable fit to transmission spectra. The high $\chi^2/n$ values for these cases indicate that the retrievals are not reliable, so it may still be possible to fit the spectra well by re-determining the mass and/or radius through minimising $\chi^2/n$ by trial and error. 

Whilst the level of systematic noise expected for EChO is not yet known, we investigate the effect of increasing the noise from the photon-limited case and find that good temperature retrievals are still possible for the hot Jupiter and Neptune cases even with a factor of 4 increase, and good gas VMR retrievals can be obtained for a factor of 2 increase. Based on the conclusions of \citet{gibson11}, we consider a factor of 3 increase over the photon-noise limit to be a worst case scenario, so do not expect systematic noise to greatly alter our conclusions.

\section*{Acknowledgements}
JKB acknowledges the support of the John Fell Oxford University Press (OUP) Research Fund for this research. SA is supported by standard grant ST/G002266/2 from the Science and Technology Facilities Council (UK) and LNF is supported by a Glasstone Fellowship at the University of Oxford. We thank the anonymous reviewer for constructive suggestions that enabled us to improve this paper.

\bibliographystyle{mn2e}
\bibliography{bibliography}

\label{lastpage}
\end{document}